\documentclass[11pt]{article}

\usepackage[margin=1in]{geometry}
\usepackage{times}
\usepackage{microtype}
\usepackage{graphicx}
\usepackage{amsmath, amssymb}
\usepackage{booktabs}
\usepackage[numbers]{natbib}
\usepackage[colorlinks=true, linkcolor=blue, citecolor=blue, urlcolor=blue]{hyperref}
\usepackage{xspace}
\usepackage{subcaption}

\usepackage{placeins}
\usepackage{enumitem}
\usepackage{booktabs}
\usepackage{tabularx}

\usepackage{xcolor}
\usepackage{pifont}

\newcommand{\ef}{\mathrm{EF}}

\usepackage{hyperref}
\usepackage{fontawesome5}
\usepackage{adjustbox}

\newcommand{\assayloop}{\textsc{AssayLoop}\xspace}
\newcommand{\assayllm}{\textsc{AssayLLM}\xspace}
\newcommand{\assayformer}{\textsc{AssayFormer}\xspace}
\newcommand{\framework}{\textsc{AssayBench-Loop}\xspace}
\newcommand{\assaybench}{\textsc{AssayBench}\xspace}

\newcommand{\screenknn}{Screen-kNN\xspace}

\newcommand{\huggingface}{\raisebox{-0.4em}{\includegraphics[height=1.5em]{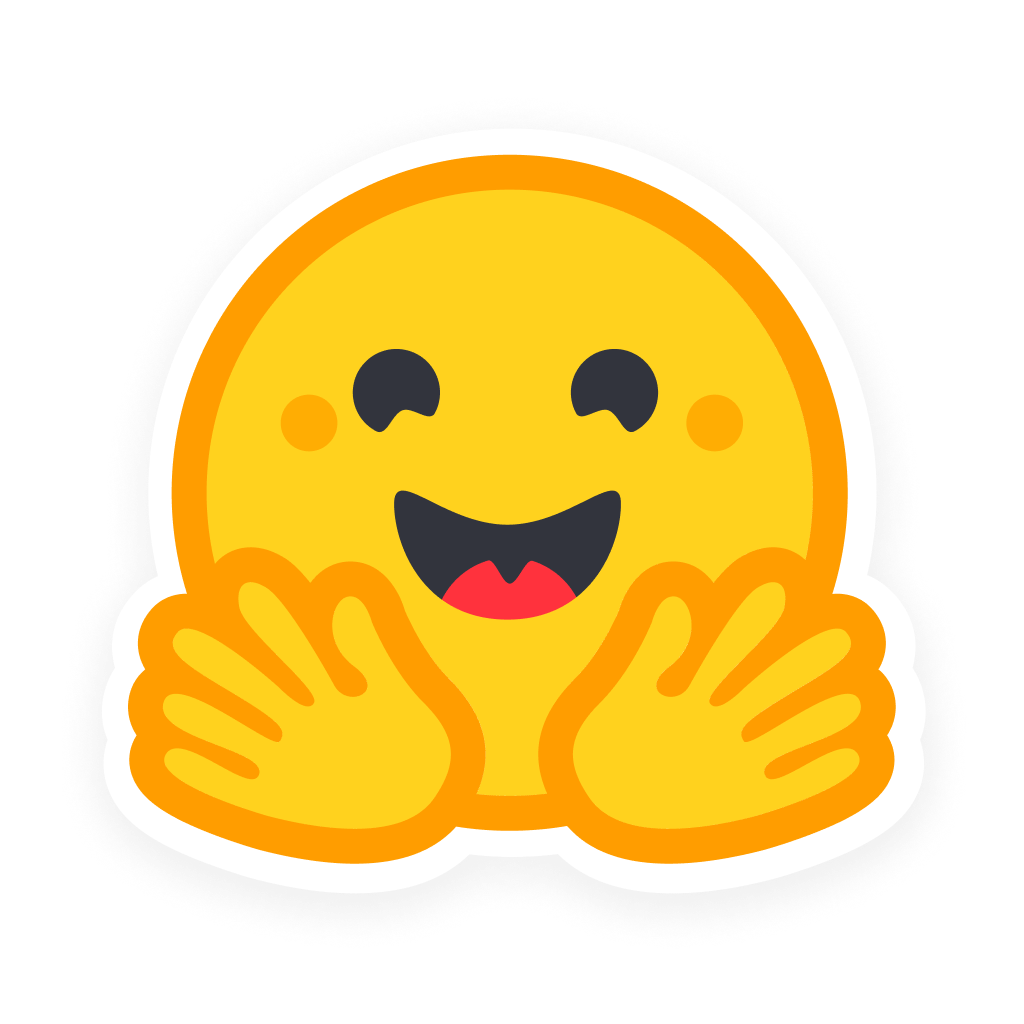}}}

\newcommand{\Rllm}{R_{\mathrm{llm}}}
\newcommand{\Cllm}{C_{\mathrm{llm}}}
\newcommand{\Fllm}{F_{\mathrm{llm}}}
\newcommand{\Rho}{R_{\mathrm{ho}}}

\usepackage{tcolorbox}
\newtcolorbox{quotebox}{
  colback=gray!10,
  colframe=gray!40,
  left=2mm,
  right=5mm,
  top=5mm,
  bottom=5mm,
  boxrule=0pt,
  leftrule=3pt
}

\usepackage{listings}
\usepackage{upquote} %
\usepackage[T1]{fontenc}

\usepackage{listings}
\usepackage{xcolor}

\usepackage{caption}

\lstdefinestyle{promptstyle}{
  basicstyle=\ttfamily\small,
  breaklines=true,
  breakatwhitespace=false,
  columns=fullflexible,
  keepspaces=true,
  frame=single,
  xleftmargin=1em,
  xrightmargin=1em,
  aboveskip=1em,
  belowskip=1em
}

\NewDocumentCommand{\gabri}
{ mO{} }{\textcolor{red}{\textsuperscript{\textit{Gabri}}\textsf{\textbf{\small[#1]}}}}

\NewDocumentCommand{\edward}
{ mO{} }{\textcolor{green}{\textsuperscript{\textit{Edward}}\textsf{\textbf{\small[#1]}}}}

\NewDocumentCommand{\carl}
{ mO{} }{\textcolor{blue}{\textsuperscript{\textit{Carl}}\textsf{\textbf{\small[#1]}}}}

\title{Biology-in-the-loop:\\ Amortized Adaptive Hit Discovery in CRISPR Screens}

\newcommand{\dnaemoji}{%
  \raisebox{-0.15em}{\includegraphics[height=1em]{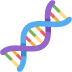}}%
}

\author{
  Carl Edwards\dnaemoji \quad Edward De Brouwer\dnaemoji \quad Xiner Li\dnaemoji \\
  Namkyeong Lee \quad Ehsan Hajiramezanali \quad Anne Biton \quad Sara Mostafavi \quad Gabriele Scalia\\
  \small{Genentech, South San Francisco, CA, USA}\\
    \texttt{\small\{{edwarc24,debroue1,lix361,scaliag\}@gene.com}}\\
  \dnaemoji \small{These authors contributed equally}
}

\date{} 

\begin{document}

\maketitle

\vspace{-10mm}
\begin{abstract}

Many biological discovery problems require experiments to be selected sequentially under constrained budgets. CRISPR screening is a prominent example, as exhaustive perturbation testing is often infeasible and candidate perturbations must instead be prioritized over multiple experimental rounds. Despite the importance of this problem, existing benchmarks for adaptive hit discovery remain limited in scale and diversity.
Here, we introduce \framework, a large-scale benchmark for adaptive hit discovery comprising 1,389 CRISPR screens across five phenotype categories. Beyond enabling systematic evaluation, its scale makes it possible to learn acquisition strategies across historical experiments.
Building on this resource, we introduce \assayloop, a sequential experimental design framework combining \assayformer, a transformer-based amortized acquisition policy trained across historical screens to adapt from experimental feedback, with LLM-derived biological priors through an adaptive handoff. In this view, completed experiments become training data for learning how accumulated evidence should guide what to test next, while LLMs provide prior biological knowledge to seed the search. We further introduce \assayllm, showing that the same principle can be extended directly to an LLM through task-specific post-training.
On temporally held-out screens, \assayloop achieves a 5.67-fold enrichment over random selection and recovers 27.7\% of hits after assaying approximately 5\% of the candidate library, outperforming existing adaptive-design methods and standalone LLMs, and \assayformer alone. Performance improves with increasing historical training data and transfers to phenotype categories excluded from training.
These results demonstrate the value of learning acquisition policies across historical experiments and combining them with broad biological priors for efficient adaptive hit discovery.

\vspace{-3mm} 
\begin{center}
    \large 
    \href{https://genentech.github.io/AssayLoop}{\faGlobe\ Website} \hspace{2em}
    \href{https://github.com/genentech/assayloop}{\faGithub\ GitHub} \hspace{2em}
    \href{https://pypi.org/project/assaybench/}{\faPython\ PyPI} \hspace{2em}
    \href{https://huggingface.co/collections/Genentech/assaybench/}{\huggingface\ Models \& Data}
    
\end{center}

\end{abstract}

\vspace{-6mm}
\begin{figure}[ht]
    \centering
    \includegraphics[width=\linewidth]{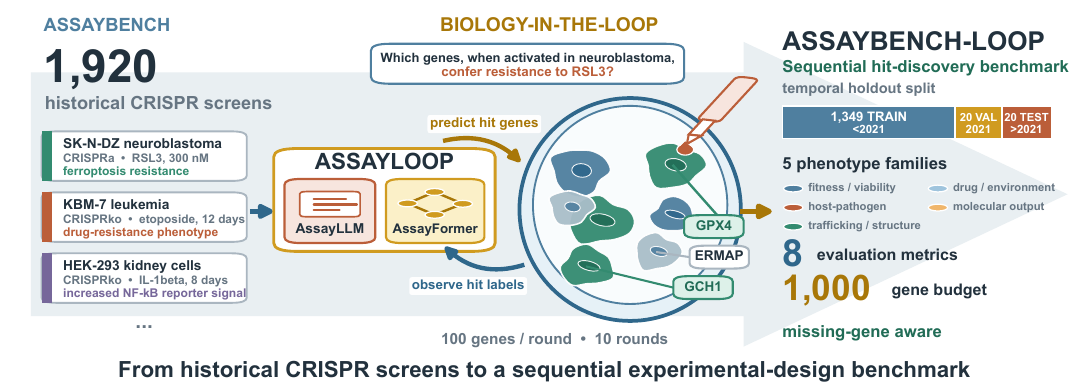}
\vspace{-6mm}
    \caption{
    Overview of \framework and \assayloop for adaptive hit discovery in CRISPR screens.
    }
    \label{fig:figure1}
\end{figure}

\section{Introduction}

CRISPR screens are central tools in functional genomics and drug discovery, enabling systematic perturbation of gene function and measurement of heterogeneous phenotypic effects~\citep{Bock2022-in,Hanna2020-pr,Przybyla2022-is}. Yet the space of possible perturbations is vast, spanning thousands of genes across diverse cellular contexts, disease models, and phenotypic readouts. Exhaustive experimentation is therefore often infeasible, particularly when the biological question requires richer models or phenotypes that cannot be measured in highly multiplexed formats. In these settings, only a fraction of candidate perturbations can be tested, often over a sequence of experimental rounds~\citep{Hanna2020-pr,Bock2022-in}. 
Efficient screen design, selecting the most informative experiments to perform next, is therefore essential for rapidly identifying biologically meaningful effects.

Sequential experimental design provides a natural framework for this setting: a model proposes experiments, observes their outcomes, and uses this feedback to determine subsequent actions under a finite budget~\citep{Rainforth2024-mm}. This model-experiment interaction forms a lab-in-the-loop process, in which experimental outcomes guide the next round of decisions. Closing this loop is a core capability for emerging AI-driven and increasingly autonomous scientific systems~\citep{202608.0273,canty2026past}.

Several methods have applied this principle to perturbation-screen design. DiscoBAX proposed an objective-driven Bayesian active-learning method encouraging coverage of diverse biological mechanisms~\citep{lyle2023discobax}. BioBO incorporates multimodal gene representations and pathway information into Bayesian optimization~\citep{li2025biobo}. Probability-of-Hit fits a probabilistic surrogate to observations collected from the current screen~\citep{Rubbi2026-oo}. %
However, these approaches treat each new experiment largely as an independent problem, fitting predictive models or acquisition strategies for each new screen. Therefore, while they can adapt to assay-specific feedback, they do not  leverage the growing collection of related experiments that have already been completed.

Large language models (LLMs) have recently been explored as an alternative, serving as acquisition policies for biological experimental design~\citep{Roohani2024-ks,Hao2025-pv,liu2024large}. By conditioning on natural-language descriptions of the experimental setting, LLMs can draw on broad scientific knowledge to prioritize plausible candidates before any observation has been collected. However, whether LLMs reliably adapt their acquisition strategies from experimental feedback remains an open question, with recent studies reaching conflicting conclusions~\citep{Gupta2025-qb,Wainrib2026-rs}. This reveals a central tension in adaptive biological experimentation: effective acquisition requires both strong prior knowledge for initial prioritization and a reliable mechanism for learning from assay-specific observations as they accumulate.

Despite the importance of adaptive hit discovery for functional genomics and target identification~\citep{van2023applications}, existing benchmarks remain limited in scale and diversity. GeneDisco introduced evaluation protocols for active-learning policies on genetic perturbation experiments~\citep{Mehrjou2021-gp}, but contains only four immunology-focused screens, and other studies have evaluated sequential design on similarly narrow collections~\citep{Roohani2024-ks,Hao2025-pv,Huang2024-ky}. No general-purpose, multi-phenotype benchmark at a scale that would support both rigorous evaluation and the training of transferable acquisition policies has been available. Such a resource is necessary not only to evaluate sequential-design methods, but to ask a broader question: can experimental systems learn from experience accumulated across many previous experiments?

Here, we address both the benchmarking gap and the methodological tension outlined above (Fig.~\ref{fig:figure1}). First, we introduce \framework, a large-scale benchmark for adaptive hit discovery comprising 1,389 CRISPR screens across five phenotype categories, equipped with dedicated metrics that account for hit-rate heterogeneity and incomplete ground truth. Beyond enabling systematic evaluation, the scale of \framework makes it possible to learn acquisition strategies from historical experiments rather than fitting each screen independently.

Building on this resource, we introduce \assayloop, a sequential experimental design framework that combines an amortized acquisition policy trained across historical screens with biological prior knowledge derived from an LLM. At its core is \assayformer, a transformer-based policy that learns, across historical assays, how experimental feedback should modify candidate prioritization, and transfers these learned strategies to new assays at inference time. In this view, completed experiments become training data for learning how to experiment, i.e., how evidence accumulated during a new campaign should guide what to test next. \assayloop complements this learned policy with LLM-derived biological priors through an adaptive handoff: an LLM guides early acquisitions when assay-specific evidence is limited, before transitioning to \assayformer as experimental observations accumulate. This design combines two complementary capabilities: broad biological prior knowledge for early prioritization, and a feedback-conditioned policy learned from historical experiments for subsequent adaptation.

We further explore whether the same principle, learning adaptive acquisition behavior from historical screens, can be realized directly within an LLM through task-specific post-training. To this end, we introduce \assayllm, a 27B-parameter LLM fine-tuned on acquisition trajectories and subsequently optimized with reinforcement learning over sequential campaigns, providing a proof of principle for this strategy.

On temporally held-out screens, \assayloop achieves a 5.67-fold enrichment over random selection and recovers 27.7\% of hits after assaying only approximately 5\% of the candidate library, outperforming existing adaptive-design methods, standalone LLMs, and its core policy used alone. We further show that the learned acquisition strategy transfers to unseen phenotype categories and improves with increasing amounts of historical training data, supporting the idea that experimental decision-making can be amortized across historical experiments. Across acquired perturbations, \assayloop retains broad biological pathway coverage while improving hit discovery. Analysis of the learned acquisition policy further reveals directional gene-gene influence patterns, including relationships supported by prior biological evidence.

Together, these results establish historical experimental repositories as a substrate for evaluating and learning transferable decision-making. 
By coupling learned policies with pretrained biological knowledge and accumulated experimental feedback, we provide a framework for improving sequential experimental design across new biological screens. More broadly, these findings suggest a path toward lab-in-the-loop systems that accumulate experimental experience across campaigns and use it to support increasingly autonomous, adaptive experimentation in biology.

\section{Results}

\subsection{A large collection of historical CRISPR screens enables learning a transferable acquisition policy for adaptive hit discovery}

We formalize adaptive hit discovery as a sequential experimental design problem (Fig.~\ref{fig:framework}A). A CRISPR screen $s$ is defined by a natural-language description $c_s$ of the experimental setup, a screen library $L_s$ of the genes measured in that screen, and binary hit labels $y_g \in \{0,1\}$ for each $g \in L_s$.
Because a single policy must act across screens with different libraries, we additionally define a fixed candidate gene pool $\mathcal{G}$, shared by all screens and containing every screen library ($L_s \subseteq \mathcal{G}$).%

At each round $t \in \{1,\ldots,T\}$, the model selects a batch $B_t$ of $b$ genes, observes their hit labels, and accumulates a history $h_t$. The goal is to maximize the total number of discovered hits within the fixed budget $T \times b$ by learning an acquisition policy $\pi(B_t \mid h_{t-1})$.

Asking whether completed experiments can inform future ones requires a collection of screens large enough to both train and evaluate acquisition strategies. To this end, we introduce \framework, a large-scale compendium of historical CRISPR screens for adaptive hit discovery built upon AssayBench~\citep{de2026assaybench}. \framework comprises 1,389 CRISPR screens across five phenotype categories, with a temporal train/validation/test split (1,349/20/20 screens; Fig.~\ref{fig:framework}B).
Validation and test screens were restricted to genome-wide assays with sufficient hit signal and non-trivial baseline recoverability, and were then selected to provide broad phenotype coverage and maximize within-category diversity of screen descriptions (Fig.~\ref{fig:framework}C; selection details in Appendix~\ref{app:dataset}, phenotype composition in Supplementary Table~\ref{tab:dataset_phenotype} and per-screen characteristics in Supplementary Table~\ref{tab:test_screens}). Test screens are substantially different from the training set (Fig.~\ref{fig:framework}E; median test-to-training hit-set Jaccard similarity $= 0.064$). At the same time, \framework spans a broad hit repertoire: although the most recurrent hits are enriched for DepMap common-essential genes, many non-essential genes also recur as hits across multiple screens (Fig.~\ref{fig:framework}F). This structure motivates acquisition policies that can exploit recurrent cross-screen structure while retaining broad coverage of the gene space, rather than collapsing onto a small set of ubiquitous hits.

\begin{center}    \includegraphics[width=\textwidth]{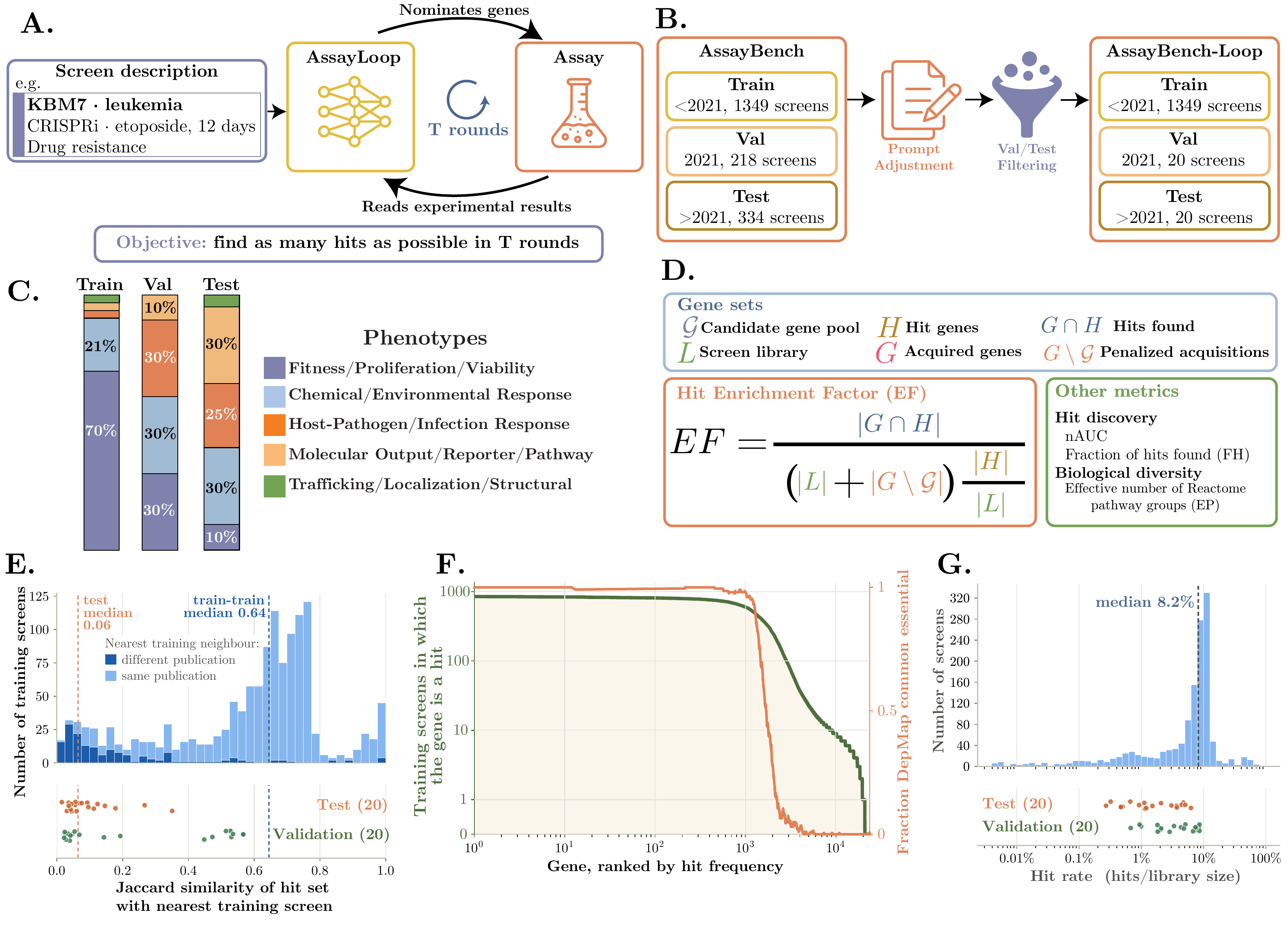}
\end{center}
\refstepcounter{figure}
\label{fig:framework}

\noindent
{\footnotesize Figure \thefigure: \textbf{\framework is a large-scale compendium and benchmark for adaptive hit discovery in CRISPR screens.}
\textbf{A}, Problem setup. An acquisition policy receives a natural-language description $c_s$ of
the screen (cell line, perturbation modality, treatment and readout) and, over $T$ rounds,
nominates a batch $B_t$ of $b$ genes to perturb, observes their binary hit labels, and conditions
the next batch on the accumulated history. The objective is to maximize the number of hits
discovered within the fixed budget $T\times b$ (here $T=10$ rounds of $b=100$ genes, i.e.\ 1,000
acquisitions per screen).
\textbf{B}, Benchmark construction. Starting from \assaybench~\citep{de2026assaybench}, screen
prompts were adapted to the sequential setting and the temporally held-out validation (2021) and
test ($>$2021) pools were filtered to 20 genome-wide screens each.
 The resulting \framework comprises 1,389 screens split temporally into 1,349 training,
20 validation and 20 test screens.
\textbf{C}, Phenotype composition of each split across the five phenotype categories. Training
screens are dominated by fitness/proliferation assays, whereas validation and test screens were
selected to spread across categories and to maximize within-category diversity of screen
descriptions.
\textbf{D}, Evaluation. Metrics are defined over the candidate gene pool $\mathcal{G}$, the screen
library $L$, the hit set $H$ and the acquired genes $G$. The primary metric is the hit enrichment
factor (EF), defined as the ratio of hits found to the number expected under random selection. The normalization penalizes hallucinated and unfilled acquisitions but does not penalize valid genes absent from the retrospective screen library. Secondary metrics include the normalized area under the
cumulative-hits curve (nAUC), the fraction of hits found (FH), and the effective number of Reactome
pathway groups (EP). \textbf{E}, Jaccard similarity between the hit set in each screen and the nearest training screen. The histogram separates within-training screens from the same or different publications; points show validation and test screens, and dashed lines mark the train-train and test medians. \textbf{F}, Training hit frequency by gene (green; symlog scale) and the fraction of DepMap common-essential genes in a 201-gene rolling window (orange). \textbf{G}, Hit-rate distribution across all 1,389 screens, with validation and test screens below (each point is a screen); the dashed line marks the median.\par}

Evaluating adaptive hit discovery requires metrics that account for substantial variation in baseline screen difficulty (Fig.~\ref{fig:framework}G) and incomplete retrospective ground truth. Not all genes in $\mathcal{G}$ are measured in every screen, and LLM-based policies can hallucinate invalid gene names or return incomplete batches. To this end, our primary metric is the hit enrichment factor (EF), defined as the ratio of discovered hits to the expected number under random selection, with a normalization that penalizes hallucinated genes but not valid genes absent from the screen library (Fig.~\ref{fig:framework}D; Methods). We additionally report the normalized area under the cumulative-hits curve (nAUC), the fraction of hits found (FH), and the effective number of Reactome pathway groups (EP) as a measure of biological diversity (eight metrics in total; Appendix~\ref{app:metrics}). All results are reported at a fixed acquisition budget per screen (10 rounds of 100 requested genes).

\framework provides a substantially larger historical training resource than existing benchmarks, while its held-out evaluation sets comprise phenotype-diverse genome-wide screens. For example, GeneDisco~\citep{Mehrjou2021-gp} is two orders of magnitude smaller, spanning four immunology screens, and other evaluations have used similarly narrow collections~\citep{Roohani2024-ks,Hao2025-pv,Huang2024-ky}. \framework provides the scale to learn a transferable acquisition strategy from historical experiments, rather than fitting each screen independently, motivating the amortized approach we describe next.

\subsection{\assayformer learns a feedback-conditioned acquisition policy across experiments, which \assayloop combines with LLM priors}

Existing sequential design methods for CRISPR screens fit a separate surrogate model for each new experiment and do not transfer acquisition strategies across screens. In contrast, we learn a shared, history-conditioned acquisition policy across all historical screens, following the amortized experimental design paradigm~\citep{Foster2021-qf,Blau2022-ri,huang2024amortized}. Here, the policy itself is amortized, meaning that a single set of parameters maps any screen description and accumulated experimental history to a ranking over candidate genes. Thus, the strategy for translating experimental feedback into the next action is learned from a large collection of completed experiments. The resulting policy transfers the learned screening strategy to new assays at inference time while adapting its decisions based on the perturbation-outcome pairs observed during the current experiment~(Fig.~\ref{fig:assayformer}A).

We instantiate this policy as a transformer encoder, \assayformer, that takes as input the screen description $c_s$ and the history of observed gene--outcome pairs $h_{t-1}$, and produces acquisition scores over all candidate genes (Fig.~\ref{fig:assayformer}B, Methods). Each observed gene is represented by a learnable embedding combined with a learned hit-indicator token. The screen description is encoded through a frozen text encoder and projected into the transformer's latent space. The transformer jointly processes the sequence of observed genes with the description token, and the output embedding at the description position is scored against all candidate gene embeddings via a bilinear head to obtain per-gene acquisition scores. At inference time, batches are selected greedily according to these scores. Full architectural details are provided in Methods.

Training proceeds in three stages (Fig.~\ref{fig:assayformer}C, Methods). First, gene embeddings are initialized using Bayesian Probabilistic Matrix Factorization (BPMF)~\citep{Salakhutdinov2008-ad} on the binary hit matrix across training screens, thereby capturing cross-screen co-hit structure.  Second, the model is trained with supervised learning to predict hit labels for unobserved genes using binary cross-entropy. To mimic intermediate states of a sequential screen, we randomly sample observed gene sets of varying sizes during training. Third, the policy is fine-tuned with reinforcement learning using group-relative policy optimization (GRPO)~\citep{shao2024deepseekmath} on full $T$-step trajectory rollouts. Notably, a context-delta reward encourages the policy to leverage the experimentally revealed history by rewarding improvements over a context-free baseline (Methods).

\assayformer is explicitly trained to use experimental feedback, but its initial acquisitions rely primarily on cross-screen structure and the limited assay-specific information contained in the screen description.
We therefore combine the learned policy with LLM-derived biological priors through an adaptive handoff strategy (Fig.~\ref{fig:assayformer}D). 

\begin{center}

    \includegraphics[width=\textwidth]{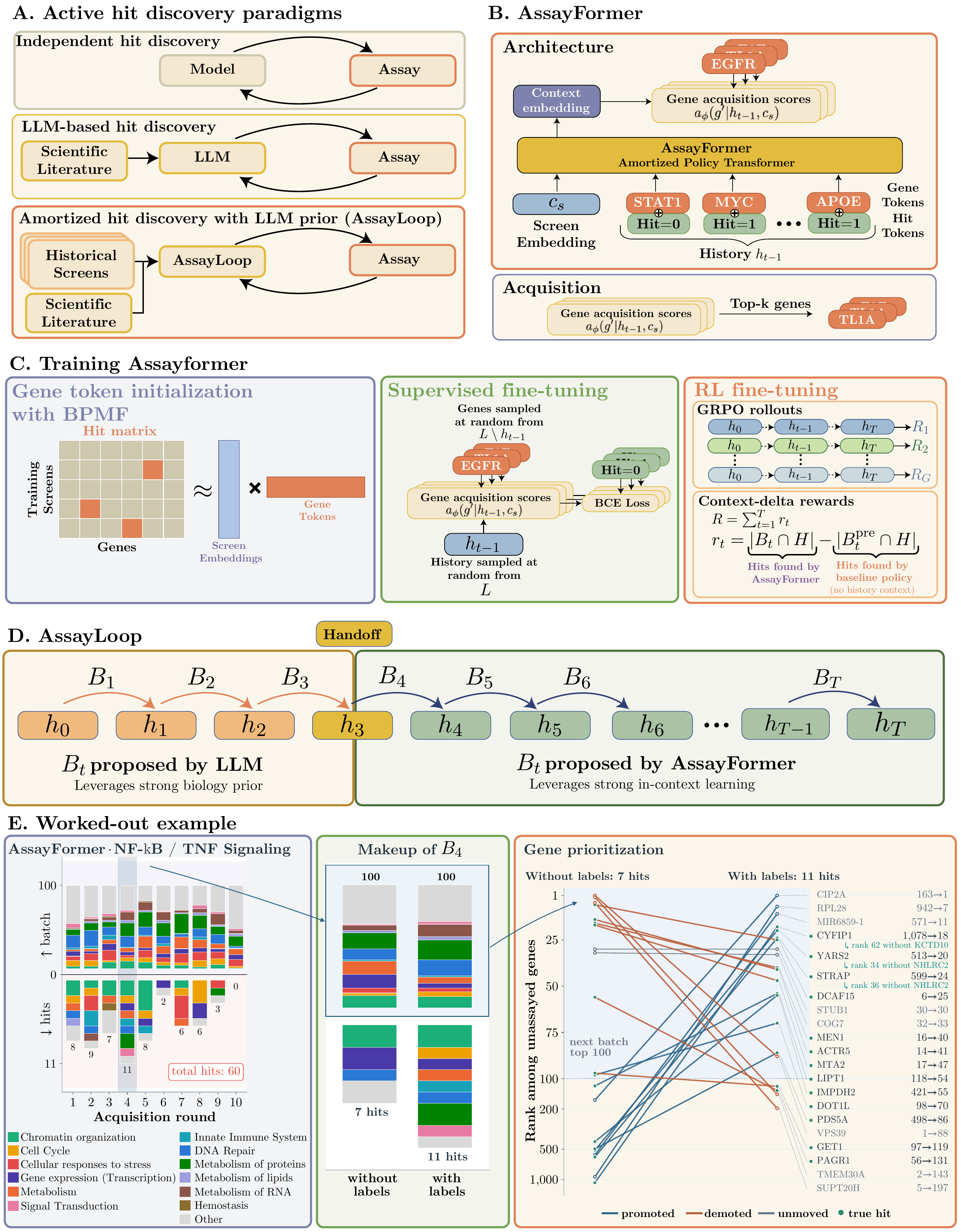}
\end{center}

\refstepcounter{figure}
\label{fig:assayformer}

\noindent
{\footnotesize Figure \thefigure: \textbf{
    \assayloop combines \assayformer, an amortized acquisition policy for adaptive hit discovery, with LLM prior knowledge.} 
\textbf{A}, Paradigms for active hit discovery. Conventional adaptive designs fit a surrogate model
independently for each new screen and do not transfer acquisition strategies across experiments.
LLM-based policies can exploit broad biological knowledge from the scientific literature, but are not explicitly trained across historical screens for feedback-conditioned adaptation. In contrast,  \assayloop follows an amortized design paradigm: it learns a shared,
history-conditioned acquisition policy across historical screens and combines it with LLM-derived priors.
\textbf{B}, \assayformer architecture. The screen description $c_s$ is embedded by a frozen text
encoder and projected into the model's latent space, while each previously assayed gene is represented by
a learnable gene token combined with a learned hit-indicator token. A transformer encoder
processes this unordered sequence, and the output embedding at the description position is scored against all
candidate gene embeddings through a bilinear head to produce per-gene acquisition scores
$a_\phi(g\mid h_{t-1},c_s)$. The next batch consists of the top-$b$ previously untested genes.
\textbf{C}, Three-stage training. Gene tokens are initialized by Bayesian probabilistic matrix
factorization (BPMF) of the binary hit matrix of the training screens.
 The policy is then trained with binary cross-entropy (BCE) to predict hit labels for unobserved genes given simulated histories of varying length, and finally fine-tuned with GRPO on full $T$-step rollouts using a context-delta reward
that credits the policy only hits recovered beyond those obtained by a context-free reference policy.
\textbf{D}, The \assayloop handoff. An LLM acts as the acquisition policy for the first $k$ rounds, providing a biologically informed warm start. \assayformer
then takes over for the remaining $T-k$ rounds and adapts to the accumulated experimental history. \textbf{E}, Worked-out example of a genome-wide CRISPR screen for regulators of TNF$\alpha$-induced NF-$\kappa$B activity in HeLa cells (169 true hits). (Left) Ten-round AssayFormer trajectory. Proposed genes are above the axis and recovered hits below, colored by broad Reactome category; round 4 is highlighted. (Middle) Composition of the round-4 batch with labels withheld or available as context: both batches contain 100 genes and recover 7 or 11 hits, respectively. (Right) Ranks of the union of the top 20 genes with and without labels as context. Blue and orange indicate at least twofold promotion or demotion when labels are supplied, gray indicates smaller changes, green points mark true hits. Teal annotations show leave-one-label-out counterfactual ranks.\par} 
\vspace{3mm}

For the first $k$ rounds ($k=3$, selected on the validation set), which we refer to as the warm-start phase, an LLM serves as the acquisition policy, processing the screen description and accumulated history in context to propose biologically informed gene selections. \assayloop then hands off to \assayformer for the remaining $T-k$ rounds, allowing the learned policy to leverage the screen-specific feedback accumulated during the warm-start phase. This design combines two complementary capabilities: biological prior knowledge for early  prioritization, provided by the LLM, and history-conditioned adaptation to subsequent experimental outcomes, learned by the trained policy.

In a representative genome-wide CRISPR screen for regulators of TNF$\alpha$-induced NF-$\kappa$B activity, (Fig.~\ref{fig:assayformer}E--G), \assayformer recovered 60 of 169 true hits within the first 1,000 assayed genes. At round 4, incorporating the 24 hit labels observed among the first 300 genes substantially reranked candidate genes for the next batch, producing a net increase from seven to eleven hits and expanding their represented pathways from four to nine categories. Leave-one-label-out counterfactuals further show that specific observations, such as KCTD10 and NHLRC2, contribute strongly to the prioritization of CYFIP1, YARS2, and STRAP (these dependencies describe the model’s decision process rather than established biological interactions).

\FloatBarrier

\subsection{\assayloop improves adaptive hit discovery on held-out screens}

\assayformer was trained on the 1,349 historical screens in the \assaybench training set and evaluated on the test set. %
A broad set of methods was also benchmarked in the same setting. These spanned adaptive experimental design algorithms, transfer learning, LLMs, and agentic approaches. Adaptive methods included Probability-of-Hit~\citep{Rubbi2026-oo} and BioBO~\citep{li2025biobo}, which fit per-screen surrogates without leveraging historical data. Transfer-based baselines included BPMF with greedy acquisition, \screenknn (which scores candidates using hit rates from similar training screens), MAML~\citep{finn2017model} (which meta-learns an initialization across training screens that can be quickly adapted on the observed hit labels), and a prior-hit frequency baseline. We evaluated several LLM families (GLM, Kimi, Claude, Qwen, Gemini, GPT), spanning open- and closed-weight models at different sizes, and test each both with and without access to feedback labels. After each acquisition round, the accumulated experimental history is appended to the next prompt.
Finally, we included two agentic harnesses specifically designed for adaptive hit discovery, LLMNN~\citep{Gupta2025-qb} and ICBR-EF~\citep{Wainrib2026-rs}, and a general coding agent based on Claude Haiku-4.5 with direct access to the training screens. Full details of all tested methods are provided in Methods. Table~\ref{tab:baselines_results_main} summarizes test set results, with full results reported in Supplementary Table~\ref{tab:baselines_results}. 

\begin{table*}[p]
\centering
\caption{\textbf{Performance comparison on held-out test screens}. Metrics evaluate enrichment factor (EF; hit enrichment relative to random selection), normalized area under the cumulative-hits curve (nAUC), fraction of hits found (FH), shortfall (SF), percentage of recovered hits that are DepMap common-essential genes (\%ess), and the effective number of Reactome level-2 pathway groups at batch (EP-B), screen (EP-S), and dataset (EP-D) scope. For EP-B/EP-S/EP-D, each gene is assigned to one of its pathway groups at random and each scope is subsampled to a fixed annotated-gene count (30, 200, and 6000, respectively), so the three metrics are not directly comparable across scopes. Random is an instantiation of the random policy, not an expected value. All performance metrics are computed separately for each screen and then averaged equally across the 20 test screens. See full results in Supplementary Table~\ref{tab:baselines_results}.}
\label{tab:baselines_results_main}
\setlength{\tabcolsep}{4pt}
\renewcommand{\arraystretch}{0.95}
\small
\begin{adjustbox}{max width=\textwidth}
\begin{tabular}{@{}lccc|cc|ccc@{}}
\toprule
\textbf{Method} & \textbf{EF} ($\uparrow$) & \textbf{nAUC (\%)}($\uparrow$) & \textbf{FH (\%)}($\uparrow$) & \textbf{SF (\%)}($\downarrow$) & $\%_{\text{ess}}$ (\textbf{\%}) & \textbf{EP-B} & \textbf{EP-S} & \textbf{EP-D} \\ \midrule

\multicolumn{9}{@{}l}{\textit{\textbf{Base LLMs}}} \\ \addlinespace
GLM-5.1 & $4.00$ & $16.3$ & $21.0$ & $10.6$ & $28.9$ & $13.3$ & $30.8$ & $64.0$ \\
\quad w/o labels & $3.65$ & $15.0$ & $19.1$ & $7.3$ & $27.6$ & $14.4$ & $33.4$ & $64.6$ \\
Gemini-3.1-Pro & $4.71$ & $20.6$ & $24.6$ & $3.6$ & $30.6$ & $14.6$ & $34.6$ & $66.2$ \\
\quad w/o labels & $4.38$ & $18.5$ & $22.9$ & $3.1$ & $29.5$ & $14.3$ & $34.3$ & $63.5$ \\
GPT-5.6 Sol & $4.81$ & $20.5$ & $25.2$ & $5.0$ & $29.6$ & $13.2$ & $29.5$ & $61.8$ \\

\midrule
\multicolumn{9}{@{}l}{\textit{\textbf{AssayLLM (Ours)}}} \\ \addlinespace
Qwen3.6-27B (base) & $2.65$ & $10.5$ & $13.9$ & $23.4$ & $30.9$ & $14.3$ & $37.5$ & $68.9$ \\
\quad + SFT (GLM-5.1 traces) & $3.56$ & $14.7$ & $18.7$ & $19.0$ & $35.9$ & $13.6$ & $34.3$ & $64.1$ \\
\quad + SFT + GRPO (= AssayLLM) & $3.69$ & $15.5$ & $19.3$ & $16.7$ & $37.8$ & $13.7$ & $34.7$ & $64.8$ \\

\midrule
\multicolumn{9}{@{}l}{\textit{\textbf{Adaptive Experimental Design Methods}}} \\ \addlinespace
Random & $1.04$ & $3.4$ & $4.9$ & $10.0$ & $18.6$ & $21.7$ & $56.9$ & $82.6$ \\
Prior hit baseline & $2.87$ & $11.5$ & $14.9$ & $1.5$ & $99.1$ & $17.8$ & $38.4$ & $48.4$ \\
\screenknn & $3.40$ & $12.5$ & $17.3$ & $3.7$ & $77.1$ & $18.9$ & $42.8$ & $54.8$ \\
BPMF~\citep{Salakhutdinov2008-ad} & $4.49$ & $12.3$ & $19.7$ & $18.0$ & $36.1$ & $20.1$ & $50.7$ & $72.5$ \\
MAML (w/ BPMF embs.)~\citep{finn2017model,Salakhutdinov2008-ad} & $3.77$ & $14.3$ & $18.6$ & $6.4$ & $74.9$ & $19.4$ & $44.4$ & $57.8$ \\
BioBO~\citep{li2025biobo} & $2.59$ & $5.9$ & $12.2$ & $10.3$ & $38.4$ & $19.9$ & $50.6$ & $72.6$ \\
Probability-of-hit~\citep{Rubbi2026-oo} & $2.41$ & $7.3$ & $11.8$ & $9.0$ & $35.2$ & $20.3$ & $53.4$ & $79.0$ \\

\midrule
\multicolumn{9}{@{}l}{\textit{\textbf{Agent Harnesses}}} \\ \addlinespace
Haiku-4.5 Agent & $3.20$ & $14.3$ & $16.8$ & $20.6$ & $49.8$ & $18.6$ & $47.3$ & $68.3$ \\
LLMNN~\citep{Gupta2025-qb} & $2.39$ & $9.4$ & $12.5$ & $0.1$ & $22.6$ & $20.3$ & $52.1$ & $78.6$ \\
ICBR-EF~\citep{Wainrib2026-rs} & $2.76$ & $9.3$ & $14.5$ & $0.9$ & $32.2$ & $18.9$ & $44.1$ & $68.5$ \\

\midrule
\multicolumn{9}{@{}l}{\textit{\textbf{AssayFormer (Ours)}}} \\ \addlinespace
AssayFormer (random embs.) & $2.47$ & $9.2$ & $12.3$ & $5.4$ & $86.1$ & $18.5$ & $40.0$ & $51.5$ \\
\quad + BPMF Embeddings & $3.83$ & $13.2$ & $18.3$ & $10.2$ & $37.5$ & $18.3$ & $42.7$ & $56.9$ \\
\quad + GRPO (= AssayFormer) & $4.83$ & $17.2$ & $23.2$ & $10.0$ & $40.7$ & $19.3$ & $46.7$ & $65.0$ \\

\midrule
\multicolumn{9}{@{}l}{\textit{\textbf{AssayLoop (Ours)}}} \\ \addlinespace
GLM-5.1 $\rightarrow$ AssayFormer & $5.27$ & $19.8$ & $25.5$ & $8.5$ & $31.6$ & $18.8$ & $47.5$ & $71.0$ \\
Gemini-3.1-Pro $\rightarrow$ AssayFormer & $5.67$ & $21.7$ & $27.7$ & $7.6$ & $30.1$ & $18.4$ & $46.6$ & $70.4$ \\
GPT-5.6 Sol $\rightarrow$ AssayFormer & $5.67$ & $21.7$ & $27.6$ & $7.8$ & $29.4$ & $18.1$ & $45.9$ & $69.8$ \\
AssayLLM $\rightarrow$ AssayFormer & $5.05$ & $18.6$ & $24.7$ & $7.8$ & $34.8$ & $18.8$ & $47.5$ & $70.8$ \\
Handoff-trained AssayLLM → AssayFormer & $5.23$ & $18.7$ & $25.1$ & $9.1$ & $39.2$ & $18.9$ & $47.0$ & $68.3$ \\

\bottomrule
\end{tabular}
\end{adjustbox}

\end{table*}

Overall, \assayloop achieves the strongest performance, with both Gemini-3.1-Pro and GPT-5.6~Sol warm starts reaching EF $\approx$ 5.67, nAUC 21.7\%, and recovering approximately 27.7\% of hits at ~5\% effective library coverage.
This exceeds standalone Gemini-3.1-Pro (EF $4.71$) and GPT-5.6 Sol (EF $4.81$), as well as \assayformer alone (EF $4.83$), supporting the complementarity of the LLM warm start and \assayformer feedback-conditioned policy. 

Analyzing standalone LLM policies, we observe relatively strong performance, led by GPT-5.6 Sol, Gemini-3.1-Pro, and GLM-5.1. Removing outcome labels consistently reduces performance, indicating that LLMs use experimental feedback to inform subsequent acquisitions. However, the magnitude of the reduction is modest relative to the performance retained without labels, suggesting that their acquisition behavior remains strongly driven by information available before observing assay-specific outcomes. These results help revisit and contextualize previous findings on LLM-driven sequential experimental design~\citep{Gupta2025-qb,Wainrib2026-rs} (additional details in Appendix \ref{app:feedback}), while highlighting a limitation of standalone LLM policies that \assayloop is explicitly designed to overcome.

\assayloop achieves considerably higher mean EF than methods that adapt within a screen without leveraging historical screen data, including BioBO, Probability-of-Hit, LLMNN, and ICBR-EF. The comparatively stronger performance of BPMF, \screenknn, and MAML, methods not specifically designed for adaptive hit discovery but that leverage available training screens, further underscores the value of transferring knowledge from historical experiments, a central motivation behind the design of \assayformer.

Figure~\ref{fig:results}A illustrates the complementary strengths of the two components of \assayloop. The LLM (Gemini 3.1 Pro shown in the figure) performs strongly in the early rounds, where its extensive biological prior gives it an advantage over methods that must learn from iterative feedback, but its performance gradually plateaus. By contrast, \assayformer relies primarily on experimental feedback: it starts from lower scores, with limited screen-specific information before the first batch, but improves steadily as outcomes accumulate. \assayloop combines these strengths, leveraging the LLM to prioritize promising hits in the initial rounds before handing off to the learned policy, which expands discovery by adapting to the accumulating experimental feedback. Supplementary Fig.~\ref{fig:QA_composition} illustrates the same pattern in more detail for two individual test screens.

\vspace{-4mm}
\subsection{Adaptive experimental capability scales with historical experience and transfers across phenotype categories}

Having established \assayloop performance on held-out screens, we next investigated how historical screens contribute to policy learning, how far the learned policy transfers beyond the biology represented in training, and which design choices drive performance.

\begin{figure}[h!]

{\centering
\includegraphics[width=\textwidth]{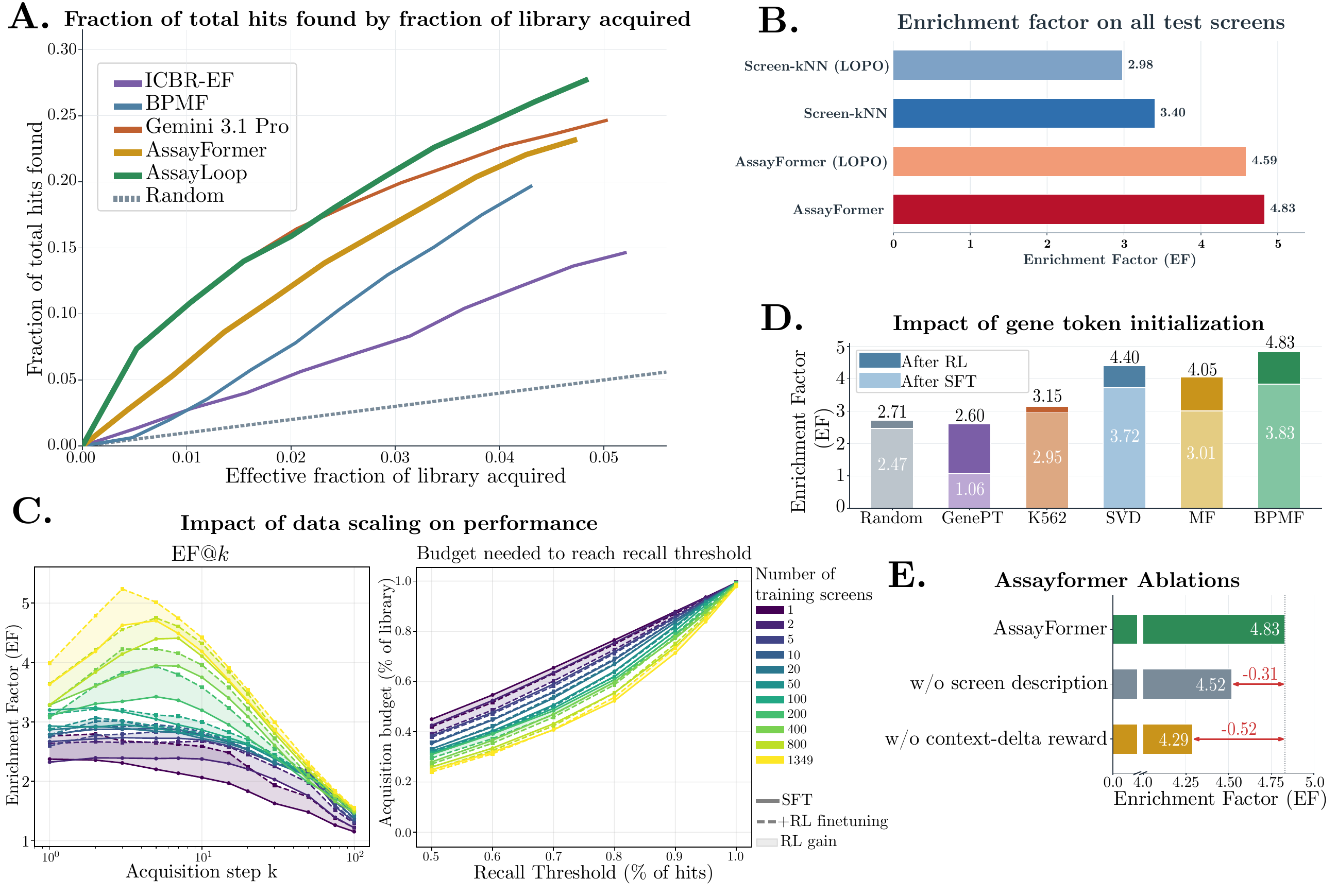}}

\refstepcounter{figure} 
\label{fig:results}

{\footnotesize Figure: \thefigure: \textbf{\assayloop outperforms existing methods at adaptive hit discovery and improves
with historical experience.} All panels report performance on the 20 temporally held-out test
screens of \framework ($T=10$ rounds of $b=100$ genes).
\textbf{A}, Fraction of all screen hits recovered as a function of the effective fraction of the
library acquired, averaged over test screens. \assayloop recovers hits fastest throughout the
campaign. Gemini 3.1 Pro is competitive over the first rounds, reflecting a strong biological prior,
but its trajectory flattens as observations accumulate, whereas \assayloop hands off to \assayformer
and continues to gain.
\textbf{B}, Phenotype-held-out transfer. EF for \assayformer and \screenknn trained on the
full training set, and in a leave-one-phenotype-out (LOPO) regime in which every training screen
belonging to the evaluated phenotype category is removed. \screenknn degrades sharply under LOPO,
while \assayformer is largely preserved and, under LOPO, still exceeds \screenknn trained on all
phenotypes.
\textbf{C}, Data scaling. 216 \assayformer models were trained on nested subsets of 1 to 1,349 
training screens (more models were trained in the low-data regime to reduce noise). Left, EF at acquisition step $k$; right, the acquisition budget (as a percentage of
the library) required to reach a given recall of the screen hits. Solid lines, supervised
fine-tuning; dashed lines, after GRPO fine-tuning; shading, the gain attributable to RL. Performance
increases monotonically with the number of training screens, with no clear plateau over the evaluated range. For this scaling analysis, we omit the screen-description input to isolate scaling trends of the in-context (i.e., history-conditioned) capabilities.
\textbf{D}, Gene token initialization. EF after SFT (light) and after RL fine-tuning (dark)
for gene embeddings initialized at random, from GenePT, from K562 expression, and from three
factorizations of the training hit matrix (SVD, MF, BPMF). Initializations that encode cross-screen
hit structure substantially outperform embeddings derived from external biological data.
\textbf{E}, \assayformer ablations. Removing the screen description $c_s$ from the input, or
replacing the context-delta reward with EF directly, both reduce EF.\par}

\end{figure}

First, we investigated data-scaling behavior by examining how \assayformer performance varies with the number of available training screens.
As shown in Figure~\ref{fig:results}C, mean performance increases across the evaluated training-set sizes, with EF@10 more than doubling over the scaling range and no clear evidence of saturation at the largest dataset size.  %
By contrast, increasing model capacity yields little additional benefit across a nearly 40-fold range in parameter count, indicating that performance in this regime is primarily limited by the number of historical screens rather than model size (Appendix~\ref{app:model_scaling}).

We next evaluated transfer to phenotype categories absent from training using leave-one-phenotype-out (LOPO) models across the five phenotype categories in \framework, holding out one category for evaluation while training on screens from the remaining four. The amount of data withheld varies substantially across these experiments
(Supplementary Table~\ref{tab:dataset_phenotype}), therefore we interpret LOPO primarily as a test of robustness to phenotype exclusion. Across the full test set, \assayformer trained under LOPO reaches EF $4.59$, compared with $4.83$ under full training, resulting in a smaller mean EF drop than \screenknn LOPO ($0.24$ versus $0.42$; Fig.~\ref{fig:results}B). Overall, \assayformer under LOPO remains above other methods trained on all phenotypes, including \screenknn ($3.40$), MAML ($3.77$), and BPMF ($4.49$). These results indicate that \assayformer performance is not solely dependent on access to training screens from the same phenotype category, supporting its generalization to unseen phenotype categories.

\FloatBarrier

Finally, we performed multiple ablations to identify important architectural choices (Fig.~\ref{fig:results}D-E, full ablations in Supplementary Table~\ref{tab:ablation_results}). Two design choices contributed meaningfully to performance and offered additional insight into how the policy operates. The first aspect is gene token initialization: in our experiments, embeddings derived from the training hit matrix (BPMF, SVD, MF) outperform those derived from external biological data, such as GenePT~\citep{Chen2024-zm} or K562 essential-gene Perturb-seq experiments~\citep{littman2025gene}, with BPMF resulting in the best performance after RL fine-tuning (EF $4.83$ vs.\ $4.40$ for SVD; Fig.~\ref{fig:results}D). The second aspect is that policy training benefits from \emph{both} conditioning on the assay and explicitly rewarding the policy for exploiting that context: either removing the screen description $c_s$ or replacing the context-delta reward (which explicitly encourages leveraging the context) with terminal EF degrades performance (Fig.~\ref{fig:results}E). While conditioning on the screen description generally improves model performance, we also notice that \assayloop's handoff design provides a broader mechanism for incorporating screen-specific contextual knowledge through the LLM in the early rounds, which leads to a substantial boost in performance~(Table~\ref{tab:baselines_results_main}).

\FloatBarrier

\begin{center}
    \includegraphics[width=\textwidth]{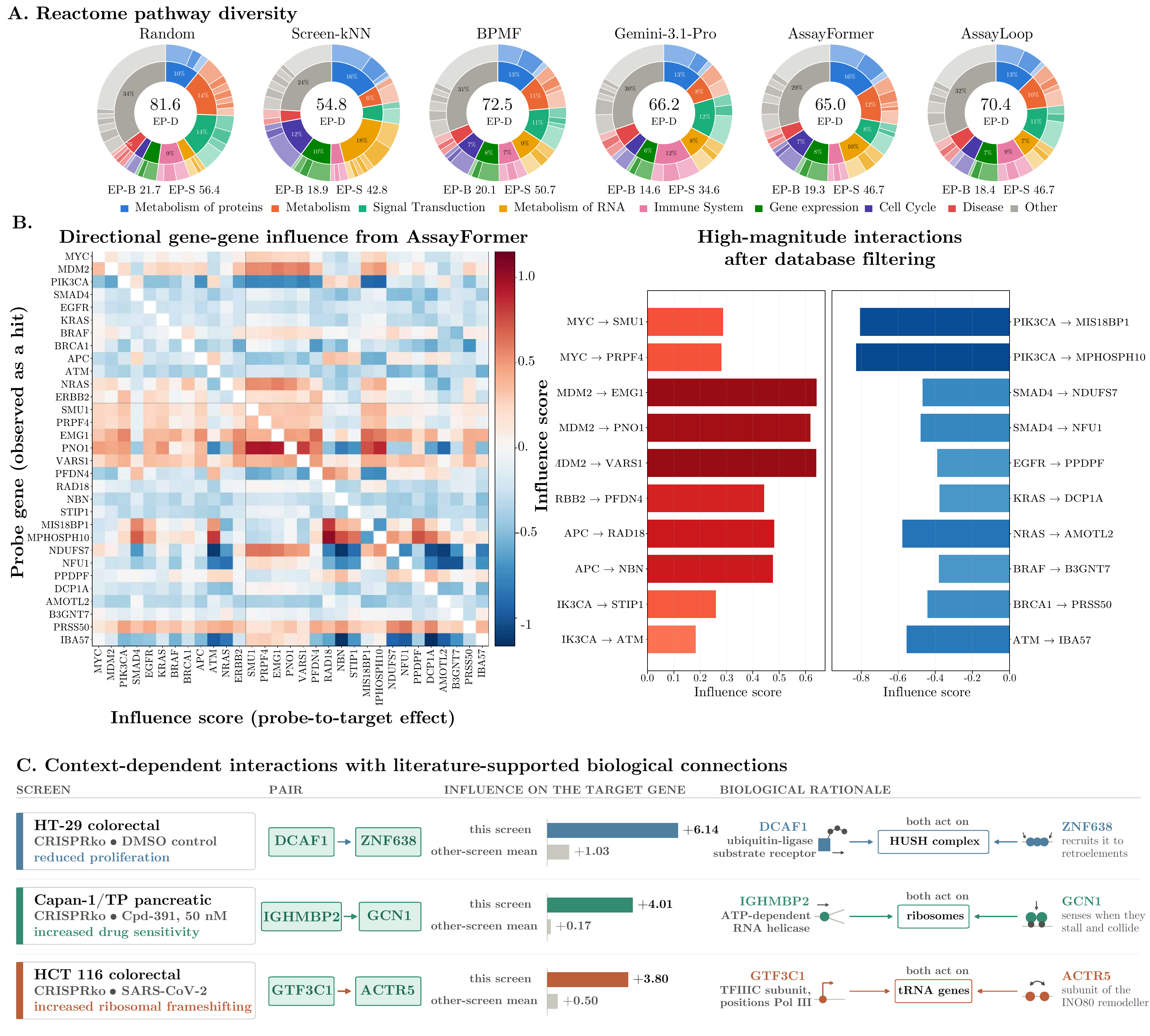}
\end{center}

\refstepcounter{figure}
\label{fig:biodiversity}

\noindent
{\footnotesize Figure \thefigure: \textbf{\assayloop acquires biologically diverse perturbations and encodes directional
gene-gene relationships.}
\textbf{A}, Reactome pathway composition of the genes acquired by six representative methods,
pooled over the 20 test screens (1,000 acquisitions per screen). Inner ring, the eight largest
top-level Reactome categories, with all remaining roots pooled into \emph{Other}; outer ring,
second-tier groups within each category, shown as tints of the parent hue. \emph{Random} is the
composition of the annotated gene universe, i.e.\ the expected composition of a uniformly drawn
batch (EP metrics are calculated for the expected distribution). The number at the centre is the dataset-level effective number of pathway groups (EP-D), values below each ring give the corresponding
batch-level (EP-B) and screen-level (EP-S) quantities. \screenknn is the most concentrated
repertoire (EP-D $54.8$), over-weighting RNA metabolism; Gemini-3.1-Pro retains
broad dataset-level breadth (EP-D $66.2$) but has lower diversity within individual batches (EP-B $14.5$); \assayloop combines high dataset-level diversity (EP-D $70.4$) with broad batch
coverage (EP-B $18.4$).
\textbf{B}, Directional gene-gene influence learned by \assayformer. Influence is defined as
the change in the acquisition score of a target gene when a probe gene is added to the observed
history as a hit, relative to a background context of 50 randomly drawn genes, averaged over 10
independently sampled backgrounds. Influence patterns are calculated under a fixed, generic screen description to inspect model-wide influence patterns learned by the model. Left, influence matrix for a panel of 31 genes comprising 12
canonical cancer drivers (upper-left block) and 19 genes from functional modules; rows are probe
genes and columns are target genes. The matrix is markedly
asymmetric, showing that the model encodes directional updates rather than gene
similarity. Right, the highest-magnitude probe$\rightarrow$target pairs remaining after removal of pairs annotated in databases (Methods).
\textbf{C}, Context-dependent influences with literature-supported biological connections. Bars compare the influence of each probe–target pair in the indicated focal screen with its mean influence across the other 19 held-out screens, holding all other inputs fixed. Both genes are true hits in the focal screen, and the pair is absent from databases. Cartoons summarize literature-supported shared biological contexts involving HUSH recruitment, translational stress sensing, and TFIIIC/INO80 activity at tRNA genes.
\par}

\subsection{\assayloop preserves biological breadth while enriching for hits}

Beyond hit discovery performance, we examined the biological diversity of acquired perturbations using the effective number of Reactome pathway groups (EP) at three scopes: within a batch (EP-B), within a screen (EP-S), and across the full test dataset (EP-D).

As shown in Table~\ref{tab:baselines_results_main} and Figure~\ref{fig:biodiversity}A, standalone LLMs show lower diversity within batches and screens (EP-B $13$--$15$, EP-S $30$--$35$), but substantially broader coverage across the full dataset. For example, Gemini-3.1-Pro reaches EP-D $66.2$, close to BPMF ($72.5$) and BioBO ($72.6$). In contrast, it achieves a proportionally much lower within-screen diversity (EP-S $34.6$) compared to those two methods ($50.7$ and $50.6$). Such behavior is consistent with LLMs focusing on different screen-specific biological programs across assays, while exploring a narrower biological neighborhood within each individual screen. Such focused prioritization is well suited to providing an initial warm start, but may limit exploration over multiple acquisition rounds. 
Further, we find that methods that do not leverage training screens, such as BioBO and Probability-of-Hit, maintain high diversity within screens and across the dataset, consistent with broader exploration rather than concentration on historically frequent hits.

\assayformer similarly explores more broadly within individual batches and screens (EP-B $19.3$; EP-S $46.7$), while maintaining high dataset-level diversity (EP-D $65.0$). \assayloop retains comparable breadth while improving hit-discovery performance: with Gemini-3.1-Pro, it achieves similar or higher batch-, screen and dataset-level diversity (EP-B $18.4$ vs. $19.3$; EP-S $46.6$ vs. $46.7$; EP-D $70.4$ vs. $65.0$), while increasing EF from 4.83 to 5.67. Thus, \assayloop combines the focused early prioritization of the LLM with \assayformer biological breadth across the screen, while achieving higher hit enrichment than either component alone.
Additional per-model diversity analysis, including pathway composition across LLM families, is provided in Appendix~\ref{app:diversity}.

\subsection{\assayformer encodes directional gene–gene influence patterns}

Because \assayformer updates its acquisition scores as observations accumulate, we can ask how observing one gene as a hit changes the prioritization of others. We define the
\emph{influence} of a probe (that is, observed) gene on a target gene as the change in the target gene's acquisition score when the probe is added to an otherwise matched history as a hit, measured
against a background context of 50 randomly drawn genes and averaged over 10
independently sampled contexts (Methods). The influence therefore captures a directional update encoded by the learned acquisition policy, rather than a calibrated interaction probability or evidence of causality. Nonetheless, these learned dependencies provide a way to interrogate the biological structure underlying the learned policy and to nominate gene relationships for further investigation.

First, we probed model-wide influence patterns under a fixed generic screen description, averaging over randomized background histories (Methods). Figure~\ref{fig:biodiversity}B (left) shows the influence matrix for a panel of 31
genes: 12 canonical cancer drivers and 19 genes drawn from functional modules. Within this panel, the learned relationships are asymmetric: $44\%$ of reciprocal pairs have influences of
opposite sign. For example, adding MDM2 to the history as a hit increases the predicted score of
PFDN4 by $0.33$, whereas adding PFDN4 decreases the predicted score of MDM2 by
$0.46$. These asymmetric relationships cannot be explained by a purely symmetric gene-similarity representation: observing one gene can change the prioritization of another in a direction that is not reciprocated.

To ask whether these learned relationships extend beyond established annotations, we removed pairs represented in STRING, CORUM, SIGNOR, MSigDB, or
Reactome~\citep{szklarczyk2023string,giurgiu2019corum,lo2023signor,subramanian2005gene,milacic2024reactome} and examined the highest-magnitude remaining influences
(Fig.~\ref{fig:biodiversity}B, right).
Among the positive pairs, including MYC in the hit history increases the score of SMU1 and PRPF4. These learned associations are consistent with established links between oncogenic MYC and dependence on spliceosome machinery~\citep{hsu2015spliceosome,koh2015myc,ciesla2021}. 
Similarly, MDM2 increases the prioritization of EMG1 and PNO1, two factors essential for small-subunit ribosome biogenesis. Biologically, disrupting ribosome biogenesis stabilizes p53 via the RPL5/RPL11-5S-RNP-MDM2 nucleolar surveillance pathway~\citep{hannan2022nuclear,castillo2023structure}. 
Supporting a functional connection, PNO1 depletion has been shown to increase RPL11–MDM2 association and stabilize p53 in TP53-wild-type colorectal cancer cells~\citep{shen2019ebf1}.
Among negative influences, several high-magnitude pairs have plausible pathway-level precedent but lack direct experimental validation. For example, SMAD4 decreases the prioritization of NDUFS7 and NFU1, consistent with reported links between SMAD4 loss and altered mitochondrial respiration in pancreatic cancer~\citep{ezrova2021smad4}.
Similarly, the NRAS-to-AMOTL2 influence is consistent with indirect pathway-level evidence: although oncogenic NRAS activates Hippo signaling in melanocytes, and parallel BRAF-driven MAPK activation reduces expression of the YAP/TAZ target AMOTL2~\citep{vittoria2022hippo}, AMOTL2 was not directly evaluated under NRAS mutation.

Because \assayformer conditions jointly on the screen description and observed history, we next explore whether learned influences vary across assay contexts. We search for pairs with substantially stronger influence in one held-out screen than across the remaining 19, requiring both genes to be hits in the focal screen and excluding pairs annotated in databases (Methods). For input context, we use the first 100 genes suggested by the LLM (Gemini-3.1), following the \assayloop protocol, and hit labels assigned from the ground truth data. This exploratory analysis identified three examples with literature-supported pathway-level connections (Fig. \ref{fig:biodiversity}C): DCAF1→ZNF638, IGHMBP2→GCN1, and GTF3C1→ACTR5. In each case, influence is substantially larger in the focal screen than across other test contexts. These examples illustrate context dependence in the learned policy and nominate hypotheses for follow-up, rather than establishing direct genetic interactions.

\subsection{Post-training LLMs can improve sequential hit discovery}

\begin{figure}[h!]

{\centering
\includegraphics[width=\textwidth]{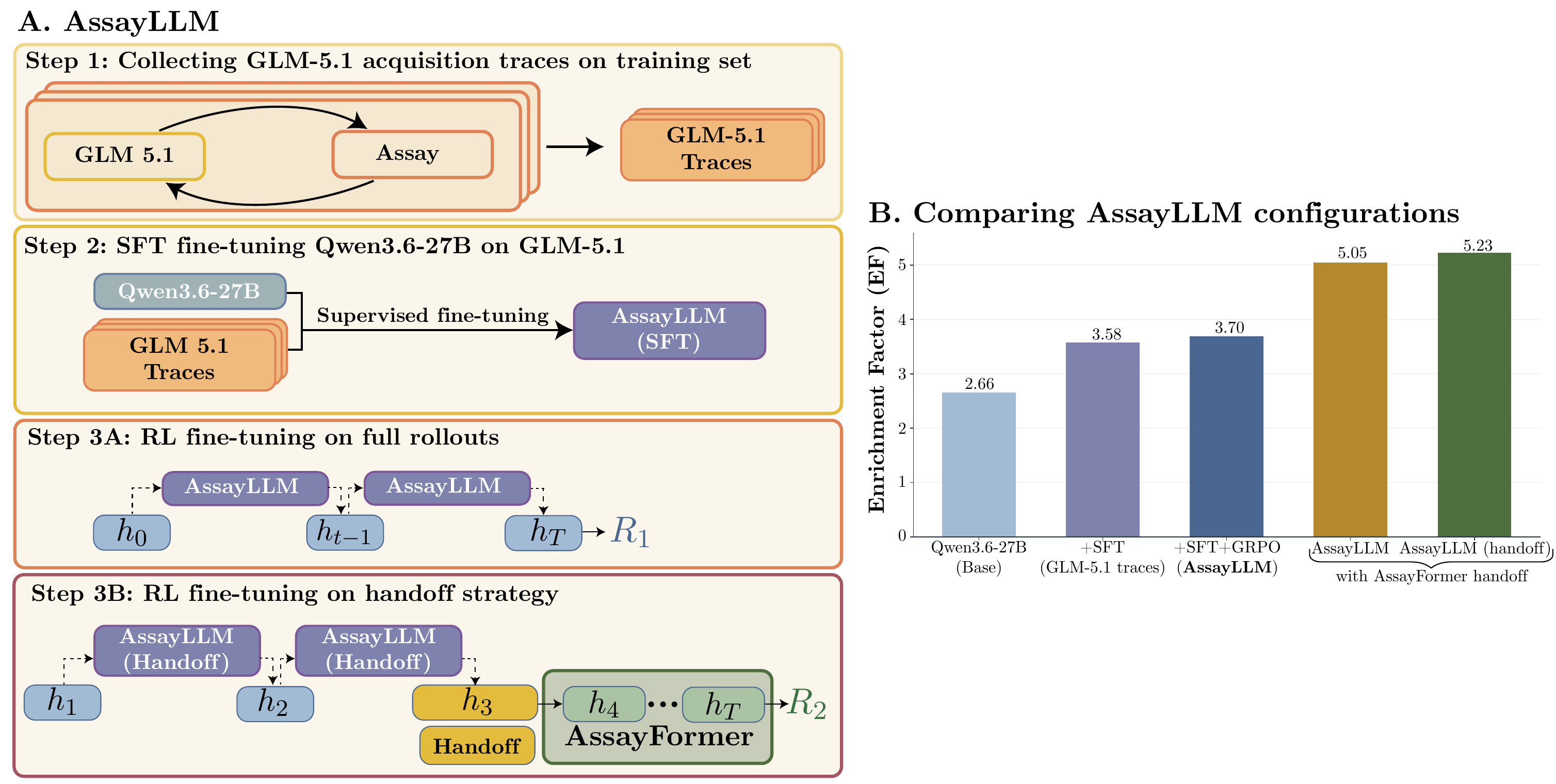}}

\refstepcounter{figure} 
\label{fig:assayllm}
\vspace{1mm}
{\footnotesize Figure \thefigure:
\textbf{Post-training improves sequential hit-discovery performance in an open-weight LLM.}
\textbf{A}, \assayllm training pipeline. Acquisition trajectories are first collected by running GLM-5.1 as an acquisition policy over the \framework training screens. A 27B-parameter Qwen3.6 backbone is then supervised fine-tuned (SFT) on these trajectories, exposing it to the sequential acquisition protocol and experimental feedback while training it to emit parseable gene batches. Two variants are optimized with group-relative policy optimization (GRPO): a standalone policy
trained on full $T$-round rollouts (3A), and a warm-start policy trained on the handoff campaign, in
which \assayllm controls the first $k=3$ rounds before the frozen \assayformer completes the
remaining rounds (3B).
\textbf{B}, Enrichment factor (EF) on the 20 test screens for the Qwen 3.6 27B base model, after SFT
on GLM-5.1 traces, and after SFT+GRPO (\assayllm), each acting as a standalone policy; and for the
two \assayloop configurations in which \assayllm provides the warm start before handing off to
\assayformer. Post-training raises standalone EF by 39\% over the base model, and the
handoff-trained variant approaches the GLM-5.1$\rightarrow$\assayformer configuration.\par}

\end{figure}

Our preceding results show that historical experiments can be used to learn a feedback-conditioned acquisition policy that generalizes to held-out screens, and that this capability complements the biological knowledge encoded in pretrained LLMs. We therefore asked whether adaptive acquisition behavior could be learned directly within an LLM through task-specific post-training, thereby directly integrating learned experimental decision-making with pretrained biological knowledge.

As a proof of concept, we introduce \assayllm, a domain-specific LLM acquisition policy built on the 27B-parameter Qwen 3.6 backbone. \assayllm is trained in two stages (Fig.~\ref{fig:assayllm}A): supervised fine-tuning (SFT) on complete acquisition trajectories generated by a stronger teacher model (GLM-5.1),  followed by group-relative policy optimization (GRPO) to directly optimize screen-level hit discovery, with rollouts from the same
screen forming a group to control for screen difficulty (Methods).

Post-training substantially improves performance on \framework (Fig. ~\ref{fig:assayllm}B; Table~\ref{tab:baselines_results_main}). Starting from Qwen3.6-27B, SFT increases EF from 2.65 to 3.56, and subsequent GRPO further increases EF to 3.69, corresponding to a 39\% improvement over the base model. \assayllm approaches its larger GLM-5.1 teacher (EF 4.0), showing that a substantial fraction of the teacher's acquisition performance can be transferred to a smaller open-weight model. The post-training improvement from EF 2.65 to 3.69 substantially narrows the gap to \assayformer (4.83). These gains, combined with the strong zero-shot acquisition performance of frontier LLMs (Table~\ref{tab:baselines_results_main}), motivate extending the same strategy to stronger pretrained LLMs as a path toward more capable standalone adaptive policies.

We further asked whether post-training could optimize an LLM specifically for its role as the warm-start policy in \assayloop. We therefore trained a second \assayllm variant in the handoff setting, in which the LLM controls the first $k$ rounds before handing the accumulated context to \assayformer (Methods). In this setting, the objective rewards not only overall hit discovery but also the quality of the experimental history constructed for \assayformer (Methods). Using \assayllm as the warm-start policy yields an EF of $5.05$, while handoff-aligned training further increases EF to $5.23$, approaching the GLM-5.1 $\to$ \assayformer configuration (EF $5.27$). Together, these results provide a proof of principle that historical screens and simulated trajectories can be used to post-train LLMs both as standalone adaptive policies and to better complement a specialized feedback-conditioned acquisition model.

\FloatBarrier
\section{Discussion}

We introduced \framework, a comprehensive benchmark for adaptive hit discovery in CRISPR screens, spanning 1,389 screens across five phenotype categories with dedicated metrics that account for incomplete libraries and LLM-specific failure modes. By providing standardized evaluation at a scale two orders of magnitude larger than previous benchmarks, \framework enables both rigorous comparison of acquisition strategies and, crucially, the training of policies that transfer across experiments.

Building on this resource, \assayloop achieves the strongest performance among the methods evaluated, recovering an average 27.7\% of hits while assaying 5\% of the candidate library.
The handoff design, in which an LLM provides biologically informed prioritization in early rounds before transitioning to \assayformer for feedback-conditioned adaptation, outperforms both components used in isolation, supporting the complementarity of these two capabilities.

Our results clarify the role of LLMs in adaptive experimental design. While we confirm that frontier LLMs do incorporate experimental feedback to some degree, their in-context adaptation remains limited compared to a policy explicitly trained for this purpose.
This finding, consistent across multiple families, suggests that current general-purpose LLMs are particularly effective as sources of biological prior knowledge, while policies explicitly trained for feedback-conditioned adaptation make more effective use of accumulating experimental outcomes.
Domain-specific post-training, as demonstrated by \assayllm, can substantially narrow this gap and improve adaptive acquisition behavior directly within an LLM: trajectory-level supervision and deployment-aligned reinforcement learning yield a 39\% improvement over the base model, and handoff-aligned training further improves its performance as the warm-start policy in \assayloop.

The data scaling analysis shows that \assayformer performance improves consistently with the number of available training screens, with no clear plateau over the range evaluated. Combined with the strong leave-one-phenotype-out generalization results, this suggests that larger and more diverse screen collections may further improve transferable acquisition policies. 

Several limitations define the scope of this study. \framework is a retrospective benchmark focused on genome-wide screens with sufficient hit signal, and the policies evaluated here remain to be validated prospectively in a live experimental loop. We also focus on binary hit discovery under a fixed acquisition budget; extending the framework to richer readouts and heterogeneous experimental costs will be important directions for future work. Finally, the retrospective formulation treats observed screen outcomes as fixed labels and does not explicitly model experimental noise across replicates.

Because \framework represents each experimental task through a natural-language description, the same benchmark formulation can in principle extend beyond the phenotypes and modalities currently represented to encompass diverse perturbation experiments and richer biological readouts. \assayllm represents a particularly promising avenue in this direction: as an LLM-based policy, it can flexibly represent new experimental readouts, perturbation types, and biological contexts through its text interface, while its post-training strategy provides a mechanism for incorporating new screening data as it becomes available. More broadly, the iterative, lab-in-the-loop paradigm explored here motivates a future in which adaptive experimental design and biological foundation model training are coupled, with each experimental campaign both advancing biological discovery and generating data that improves the models guiding subsequent experiments.

\FloatBarrier

\clearpage

\bibliographystyle{plainnat}
\bibliography{ref}

\clearpage

\section{Methods}

\subsection{Problem formulation}
\label{sec:baed}

We frame adaptive hit discovery as a special case of Bayesian adaptive experimental design (BAED)~\citep{Rainforth2024-mm}. BAED selects experiments sequentially to maximize a utility function under uncertainty. Its components are an experimental domain $D$, an observation model $p(y \mid d, \theta)$ generating outcomes for a given experiment $d$ and parametrized by an unknown parameter $\theta$ with prior $p(\theta)$, and a history $h_t = \{(d_i, y_i)\}_{i=1}^t$ of past experiment--outcome pairs. At each step, the design policy $\pi(d_{t+1} \mid h_t)$ selects the next experiment. For a fixed horizon $T$, the optimal policy maximizes

\begin{align}
    \label{eq:BAED}
    \pi^* = \operatorname*{arg\,max}_{\pi} \mathbb{E}_{\pi,p}[U(h_T,\theta)],
\end{align}
where $U(h_T,\theta)$ is a utility function over the experimental trajectory.

Let a CRISPR screen $s$ be composed of a natural-language description $c_s$ of the experimental setup, a candidate gene pool $\mathcal{G}$ of all possible genes ($|\mathcal{G}| \approx 20{,}000$), a screen library $L \subseteq \mathcal{G}$ of the genes actually measured in that screen, and binary hit labels $y_{g} \in \{0,1\}$ for each gene $g \in L$. At each round $t \in \{1,\ldots,T\}$, the model selects a batch $B_t$ of $b$ genes, observes their hit labels, and accumulates a history $h_t = \{(g, y_{g}) : g \in B_1 \cup \cdots \cup B_t\}$. The goal is to maximize the number of discovered hits within the fixed budget $T \times b$:

\begin{align}
    \label{eq:objective}
    \pi^* = \operatorname*{arg\,max}_{\pi} \mathbb{E}_{\pi}\Big[\sum_{g\in h_T} y_g\Big],
\end{align}
where the expectation is taken over full trajectories sampled from $\pi$. This corresponds to the general BAED framework of Eq.~\ref{eq:BAED} with a deterministic observation model $p(y_g \mid g) = \mathbb{I}[y=y_g]$ and utility $U(h_T)=\sum_{g\in h_T} y_g$. Throughout this work we use $T=10$ rounds of $b=100$ genes, for a total budget of 1{,}000 acquisitions per screen.

Conventional BAED methods solve each experimental-design problem independently, fitting a task-specific probabilistic model using only observations from the current experiment~\citep{Rubbi2026-oo,li2025biobo}. Although external knowledge can be incorporated through priors, kernels, or representations, the acquisition policy itself is not learned across tasks. \emph{Amortized} experimental design instead learns a shared, history-conditioned policy across a distribution of tasks, moving much of the inference cost to an offline training stage and enabling knowledge transfer to new experiments~\citep{Foster2021-qf,Blau2022-ri,huang2024amortized}. In our setting, completed large-scale CRISPR screens provide a collection of related experimental tasks from which sequential histories can be constructed retrospectively, allowing us to train an amortized acquisition policy directly from existing datasets without requiring a separate generative model.

\subsection{Dataset and benchmark construction}
\label{sec:dataset}

\framework builds upon AssayBench~\citep{de2026assaybench}, a public benchmark containing 1,920 CRISPR screens. We adopt the same temporal train/validation/test split strategy and retain all 1,349 training screens. We use its temporal split: training screens before 2021, validation screens from 2021, and test screens after 2021. Because evaluating sequential design over multiple rounds is computationally intensive, we limit the validation and test sets to 20 screens each. These screens were selected to satisfy the following criteria: genome-wide coverage ($18{,}000$--$22{,}000$ genes), at least 50 hits with hit rate below 15\%, a non-trivial LLM performance floor (AnDCG@100 $\geq 0.05$ for the best LLM from AssayBench, Gemini 3.1 Pro), phenotype distribution allocated based on the distribution of eligible test set screens, with a soft cap of six screens per category, and maximized within-phenotype diversity based on TF-IDF similarity of screen descriptions. The complete dataset composition is provided in Appendix~\ref{app:dataset}.

\subsection{Evaluation metrics}
\label{sec:metrics}

We let $\mathcal{G}$ denote the set of all possible genes, $L$ the set of genes in the gene library of a given screen (for which a label $y_g$ is available), $H$ the set of hits in $L$, and $G_T$ the set of genes acquired by the model over the $T$ rounds. Full metric definitions and implementation details are provided in Appendix~\ref{app:metrics}.

\paragraph{Hit enrichment factor (EF).} Our primary metric measures the ratio between the number of hits found and the expected number of hits under random selection. We write $N_{L} = |G_T \cap L|$ for the number of acquired in-library genes, $N_{\bar{\mathcal{G}}} = |G_T \setminus \mathcal{G}|$ for the number of hallucinated acquired genes that do not represent valid genes, and $N_{\mathrm{miss}}$ for the total
number of unfilled acquisition slots over the $T$ rounds. Finally, $h_{\text{rand}}=|H|/|L|$ denotes the random hit rate. We define

\begin{align}
    \label{eq:enrichment_factor}
    \mathrm{EF}=\frac{|G_T\cap H|}{(N_L + N_{\bar{\mathcal{G}}} + N_\text{miss})\cdot h_{\text{rand}}}.
\end{align}

This differs from the classical enrichment factor ($|G_T\cap H| / (|G_T| \cdot h_{\text{rand}})$) in the denominator: normalizing by $(N_L + N_{\bar{\mathcal{G}}} + N_\text{miss})$ does not penalize acquisition of valid genes absent from the screen library but does penalize hallucinated genes and unfilled acquisition slots.

\paragraph{Additional metrics.} We report the normalized area under the cumulative-hits curve (nAUC), which captures the full acquisition trajectory; the fraction of hits found (FH), measuring recall; the shortfall (SF), the fraction of acquisitions falling outside the screen library; and the percentage of acquired hits that are DepMap common-essential genes ($\%_{\text{ess}}$)~\citep{Tsherniak2017-fa}, which separates screen-specific biology from generic cell-death phenotypes.

We quantify the biological diversity of the acquired genes with three measures. The effective number of Reactome pathway groups (EP) is the Hill number of order 1 of the pathway distribution induced by a set of acquired genes, reported at the batch (EP-B), screen (EP-S), and dataset (EP-D) scopes~\citep{hill1973diversity}. Within a batch, the Vendi score~\citep{friedman2022vendi} is the effective number of distinct genes under a cosine-similarity kernel over GenePT embeddings~\citep{Chen2024-zm}, reported as Vendi ratio (VR), a fraction of batch size, so that higher values indicate more diverse batches. The pathway overlap (PO) is the mean pairwise Jaccard overlap of GO biological-process annotations within a batch, divided by that of a random draw from the same library: $\text{PO}=1$ matches random selection and $\text{PO}>1$ indicates genes that share annotations. Full definitions are given in Appendix~\ref{app:metrics}.

\subsection{\assayformer architecture}

\assayformer is a learned acquisition policy $\pi_\phi(B \mid h_{t-1}, c_s)$ over candidate gene batches, following the amortized design paradigm. Given a screen description $c_s$ and the current history $h_{t-1}$, it produces per-gene acquisition scores over all untested candidates.

The model is a transformer encoder with $L=3$ layers, $H=2$ attention heads, width $d_{\text{model}}=384$, feed-forward dimension $d_{ff} = 1024$, and GELU activations. The input is a sequence of tokens comprising the experiment description (DESC) and the history gene tokens: $[\text{DESC}, x_{g_1}, \dots, x_{g_{|h_{t-1}|}}]$. No positional embeddings are applied, as gene order in the history is treated as irrelevant.

Each history gene token $x_g$ combines a learnable gene embedding $v_{g} \in \mathbb{R}^{d_g}$ with a learned embedding $z_{y_g}$ of its binary hit label $y_g\in\{0,1\}$. Separate embeddings $z_0$ and $z_1$ are learned for non-hits and hits, respectively. The two are concatenated and projected to the transformer latent space:

$$ x_g = W^T\big(v_{g} \oplus z_{y_g}\big),$$

where $W\in \mathbb{R}^{(d_g + d_h) \times d_{\text{model}}}$ is a learnable projection matrix.

The screen description $c_s$ is processed by first embedding it with a frozen text encoder (OpenAI \texttt{text-embedding-3-small}), yielding $e_{\text{desc}}\in \mathbb{R}^{d_{\text{text}}}$, and then linearly projecting it into the latent space: $\text{DESC} = W_d \, e_{\text{desc}}$, where $W_d \in \mathbb{R}^{d_{\text{model}} \times d_{\text{text}}}$.

The pooled output of the DESC token through the transformer is linearly projected to the gene space to produce the screen latent $\hat{u}(h_{t-1},c_s) \in \mathbb{R}^{d_g}$. Each candidate gene $g$ is scored via a bilinear head:

$$
a_\phi(g \mid h_{t-1}, c_s) \;=\; \hat{u}(h_{t-1},c_s)^\top v_g \;+\; b_g,
$$

where $b_g$ is a per-gene bias. At inference time, the next batch $B_t$ is constructed by taking the top-$b$ previously untested genes according to $a_\phi$.

\subsection{Training procedure}
\label{sec:training}

Training proceeds in three stages: gene embedding initialization, supervised fine-tuning, and reinforcement learning fine-tuning.

\paragraph{Gene embedding initialization.}
We initialize the gene embeddings $v_{g} \in \mathbb{R}^{d_g}$ ($d_g = 10$) using a probit version of Bayesian Probabilistic Matrix Factorization (BPMF)~\citep{Salakhutdinov2008-ad, albert1993bayesian}. Let $Y_{ij}\in\{0,1\}$ indicate whether gene $j$ scored as a hit in training screen $i$. We model

\begin{equation}
  Z_{ij} = U_i^{\top} V_{g_j} + \varepsilon_{ij},\quad
  \varepsilon_{ij}\sim\mathcal{N}(0,1),
  \qquad
  Y_{ij} = \mathbb{I}\left[Z_{ij} > 0\right],
\end{equation}

with isotropic Gaussian priors $U_i\sim\mathcal{N}(0,\sigma_u^2 I_{d_g})$ on screen factors and $V_{g_j}\sim\mathcal{N}(0,\sigma_v^2 I_{d_g})$ on gene factors ($\sigma_u=\sigma_v=1$). Posterior distributions are estimated using Gibbs sampling with 2{,}000 iterations (1{,}000 burn-in, thinned by 2). Gene embeddings are set to the posterior mean $v_g = \mathbb{E}[V_g \mid Y]$. Genes absent from the factorization receive $v_g = \mathbf{0}$ and a static bias at their marginal hit frequency. Algorithmic details are provided in Appendix~\ref{app:bpmf}. This initialization is important for downstream performance, as discussed in Appendix~\ref{sec:emb_init}.

\paragraph{Supervised fine-tuning.}
The model is trained to predict hit labels for unobserved genes given simulated histories of varying length, using binary cross-entropy. For each training screen, a random subset $\mathcal{O}$ of the screen's genes (where $|\mathcal{O}| \sim \mathcal{U}\{0, 1024\}$) is revealed as observed context. To ensure the model encounters informative histories, with probability 25\% we resample $\mathcal{O}$ to contain at least one hit: we sample $n_{\text{hits}} \sim \mathcal{U}\{1, \min(|H|-1, |\mathcal{O}|)\}$ and replace $n_{\text{hits}}$ genes in $\mathcal{O}$ with randomly drawn hits, capping at $|H|-1$ to leave at least one hit unobserved. The model is trained on a balanced set $\bar{\mathcal{O}}$ of unobserved genes (all unobserved hits plus an equal number of randomly sampled non-hits):

$$
\mathcal{L}_{\text{SFT}} \;=\; \frac{1}{|\bar{\mathcal{O}}|}\sum_{g \in \bar{\mathcal{O}}}
\mathrm{BCE}\!\Big(\sigma\big(a_\phi(g \mid h_{t-1}, c_s)\big),\, y_g\Big).
$$

\noindent This yields a greedy policy $\pi_{\text{pre}}$ that selects the $b$ genes with highest scores at each step.

\paragraph{RL fine-tuning.}
The supervised policy is greedy and does not account for the fact that decisions at one step can influence the information gained at future steps~\citep{Rainforth2024-mm}. We therefore fine-tune the policy using GRPO~\citep{shao2024deepseekmath} on full $T$-step trajectory rollouts. For each training screen, we perform $G=8$ rollouts. Since the policy network is deterministic, rollout diversity is produced via Gumbel-top-$k$ sampling~\citep{kool2019stochastic}:

$$ B_t = \operatorname{Top\text{-}b}\big(\{ \tilde{s}_g: g \notin h_{t-1} \}\big); \qquad
\tilde{s}_g = a_\phi(g \mid h_{t-1}, c_s) + \tau \, \epsilon_g,\quad \epsilon_g \sim \text{Gumbel}(0,1),
$$

where $\tau = 1$ is the sampling temperature. Fresh noise is sampled at each acquisition step, so trajectories diverge as context accumulates.

The reward uses a ``context delta'' approach. At each step $t$, the reward is the difference in hits found by the context-conditioned policy relative to a frozen, context-free policy:

$$
r_t \;=\; \big|\{g \in B_t^{\pi} : y_g = 1\}\big| \;-\; \big|\{g \in B_t^{\pi_0} : y_g = 1\}\big|,
$$
\noindent
where $B_t^\pi$ is the batch sampled by the current policy and $B_t^{\pi_0}$ is sampled from the same set of unobserved genes by the frozen policy $\pi_0$ without context history. $\pi_0$ is initialized by the supervised fine-tuning phase. Optionally, we update the weights of $\pi_0$ from $\pi$ every $n$ epochs. This reward function is aligned with the hit-discovery objective in Eq.~\ref{eq:objective} while explicitly encouraging the model to exploit the revealed context to outperform the static prior. %
Advantages are computed as remaining returns $G_t^{(i)} = \sum_{s \geq t} r_s^{(i)}$, normalized across the group at each step. The policy is updated by the REINFORCE gradient of Plackett--Luce log-probabilities weighted by advantages, with an entropy bonus (coefficient 0.01). The fine-tuned policy $\pi_{\text{RL}}$ is sampled greedily at inference time.

\subsection{LLM handoff strategy}

We combine the biological prior knowledge of LLMs with the adaptive capabilities of \assayformer via a handoff strategy. For the first $k$ rounds, an LLM acts as the acquisition policy: given the screen description $c_s$ and observed history $h_{t-1}$ in context, it produces a list of $b$ genes, providing a biologically informed warm start. We then switch to the amortized acquisition policy $\pi_{\text{RL}}$ for the remaining $T-k$ rounds, allowing the learned policy to adapt to screen-specific feedback accumulated during the warm-start phase. We select $k=3$ on the validation set. We refer to the resulting composite policy as \assayloop.

\subsection{\assayllm}
\label{sec:assayllm}

\assayllm is an LLM acquisition policy post-trained for adaptive hit discovery. At acquisition round $t$, the model receives the screen description $c_s$ and the observed history $h_{t-1}$ as a multi-turn conversation, and returns a ranked list of $b$ genes defining the next acquisition batch $B_t$. We train \assayllm for two deployment settings: as a standalone policy controlling all $T$ acquisition rounds, and as a warm-start policy controlling the first $k$ rounds before handing the accumulated history to \assayformer.

Both variants use the 27B-parameter Qwen 3.6 backbone and share a two-stage post-training procedure. We first perform supervised fine-tuning (SFT) on acquisition trajectories and reasoning traces generated by GLM-5.1. This stage teaches the model to follow the sequential acquisition protocol, incorporate experimental feedback, avoid repeated selections, and reliably produce parseable gene batches. We then apply GRPO~\citep{shao2024deepseekmath} to directly optimize campaign-level hit-discovery outcomes. Rollouts from the same screen form a GRPO group, controlling for variation in screen difficulty.

To define the rewards, let $V_t = (B_t \cap L_s) \setminus \operatorname{dom}(h_{t-1})$ denote the distinct, previously unobserved, in-library genes returned at round $t$, and let $|H_s|$ denote the total number of hits in screen $s$. For the $k$ rounds controlled by the LLM, we define

\begin{align}
\Rllm(k)
    &=
    \frac{1}{|H_s|}
    \sum_{t=1}^{k}
    \sum_{g\in V_t} y_g,
    &
\Cllm(k)
    &=
    \frac{
        \left|\bigcup_{t=1}^{k}V_t\right|
    }{k b},
    \nonumber\\
\Fllm(k)
    &=
    \frac{1}{k}
    \sum_{t=1}^{k}
    \mathbb{I}\!\left[
        |V_t|\geq \frac{b}{2}
    \right].
\label{eq:assayllm-shared-reward-terms}
\end{align}

Here, $\Rllm$ measures the fraction of screen hits found directly by \assayllm, $\Cllm$ measures the fraction of its acquisition budget yielding distinct scoreable observations, and $\Fllm$ measures whether the model returns sufficiently complete and valid batches.

\paragraph{Standalone reward.} In the standalone setting ($k=T$), we optimize

\begin{equation}
r_{\mathrm{standalone}}
=
w_{\mathrm{llm}}\,\Rllm(T)
-
w_{\mathrm{fmt}}\left(1-\Fllm(T)\right).
\label{eq:assayllm-standalone-reward}
\end{equation}

\paragraph{Warm-start reward.} In the warm-start setting, \assayllm controls rounds $1,\ldots,k$ and the frozen \assayformer completes rounds $k+1,\ldots,T$. To measure handoff quality, we define

\begin{equation}
\Rho(k,l)
=
\frac{1}{|H_s|}
    \sum_{t=k}^{l}
    \sum_{g\in V_t} y_g,
\label{eq:assayllm-readiness}
\end{equation}

where $V_{k+1}$ is the first batch selected by \assayformer after the handoff. The warm-start reward is

\begin{equation}
\begin{split}
r_{\mathrm{warm}}
={}&
w_{\mathrm{obj}}\,\ef
+
w_{\mathrm{llm}}\,\Rllm(k)
+
w_{\mathrm{ready}}\,\Rho(k+1,T)
\\
&+
w_{\mathrm{cov}}\,\Cllm(k)
-
w_{\mathrm{fmt}}\left(1-\Fllm(k)\right).
\end{split}
\label{eq:assayllm-warm-reward}
\end{equation}

The primary term $\ef$ evaluates the complete LLM--\assayformer campaign. The $\Rllm$ term provides direct credit for hits found during the LLM-controlled rounds, while $\Rho$ credits the LLM for constructing a history that supports an effective first acquisition by \assayformer. The $\Cllm$ term encourages distinct, scoreable observations, and the format penalty discourages degenerate batches. The auxiliary terms are important because the frozen \assayformer may recover many readily discoverable hits regardless of the warm start, making the campaign outcome alone a weak learning signal. Additional implementation details are provided in Appendix~\ref{app:assayllm}.

\subsection{Gene-gene influence analysis}
\label{sec:influence}

To probe what \assayformer has learned about relationships between genes, we measure how observing one gene as a hit shifts the policy's predicted score for another. For an ordered probe--target pair $(p,g)$, we draw a synthetic background history $h_{\text{bg}}$ of $50$ genes sampled uniformly at random from the gene vocabulary, each assigned a uniform random binary hit label, and compare the acquisition score of the target under this history with its score under the same history extended by the probe observed as a hit. The influence of $p$ on $g$ is the expected difference,

\begin{equation}
\label{eq:influence}
I(p \to g) \;=\; \mathbb{E}_{h_{\text{bg}}}\Big[\,a_\phi\big(g \mid h_{\text{bg}} \cup \{(p,1)\},\, c_s\big) \;-\; a_\phi\big(g \mid h_{\text{bg}},\, c_s\big)\Big],
\end{equation}

estimated by averaging over $10$ independently sampled backgrounds $h_{bg}$ with fixed random seeds. Because the per-gene bias $b_g$ is shared by both terms, the difference reduces to $\big(\hat{u}(h_{\text{bg}} \cup \{(p,1)\}, c_s) - \hat{u}(h_{\text{bg}}, c_s)\big)^\top v_g$, that is, the projection of the probe-induced shift in the screen latent onto the target gene embedding. Influence is therefore directional by construction: $I(p \to g)$ and $I(g \to p)$ are separate quantities and need not agree in magnitude or sign.

Randomizing the hit labels of the background genes prevents the measurement from being dominated by a single hypothetical screen outcome, and the shared background across the two terms means that only the probe observation differs between them. All influences are computed with the RL-fine-tuned \assayformer checkpoint under a generic screen description (``A genome-wide CRISPR knockout screen to identify essential genes''), so that the reported relationships reflect structure learned across training screens rather than a single assay context. Probe or target genes absent from the model's gene vocabulary are excluded.

\paragraph{Screen-specific influences} For computing screen-specific influences (Fig. \ref{fig:biodiversity}C), given a focal screen, we replace the generic screen description with the corresponding screen description $c_s$ and set the background context $h_{bg}$ to the first 100 genes selected by Gemini-3.1-Pro (given access to $c_s$ only), together with their ground-truth hit labels.

\subsection{Baselines}

\paragraph{Adaptive experimental design methods.} We evaluate
Probability-of-Hit~\citep{Rubbi2026-oo} and BioBO~\citep{li2025biobo},
two adaptive methods designed for CRISPR hit discovery. BioBO uses a UCB
acquisition with a biological enrichment analysis prior, while
Probability-of-Hit uses an active search acquisition function. Both use
an MLP surrogate with biologically-informed gene embeddings but do not
leverage existing CRISPR screen data.

\paragraph{Transfer-based baselines.} To assess the value of learning from historical screens, we evaluate three baselines that exploit the training data: BPMF~\citep{Salakhutdinov2008-ad} with a greedy acquisition function (Appendix~\ref{app:bpmf}); \screenknn (Appendix~\ref{app:screen_knn}), which scores candidate genes using a distance-weighted combination of hit rates from the most similar training screens; and MAML~\citep{finn2017model}, which meta-learns an initialization across training screens that can be quickly adapted via gradient steps on observed hit labels. We also include a prior-hit baseline that ranks genes by their empirical hit frequency across training screens.

\paragraph{Large language models.} We evaluate several LLM families, including GLM \cite{zeng2026glm}, Kimi \cite{team2026kimi}, Claude \cite{claude4.8}, Qwen \cite{qwen3.6-27b}, Gemini \cite{Gemini_2026}, and GPT \cite{singh2025openai}. We adapt the prompt from AssayBench~\citep{de2026assaybench} to the sequential setting: after each acquisition round, the accumulated experimental history, including previously selected genes and their observed hit labels, is appended to the next prompt. An example is shown in Appendix~\ref{app:prompt_template}.

\paragraph{Agentic harnesses.} We evaluate two agentic harnesses designed for adaptive hit discovery: LLMNN~\citep{Gupta2025-qb}, which combines LLM prior knowledge with nearest-neighbor sampling, and ICBR-EF~\citep{Wainrib2026-rs}, which uses a hypothesis registry to improve in-context learning. Neither method leverages existing screen data. We therefore also evaluate a general coding agent on top of Haiku-4.5 with access to the training screens.

\appendix

\FloatBarrier
\clearpage
\setcounter{figure}{0}
\setcounter{table}{0}
\renewcommand{\thefigure}{\arabic{figure}}
\renewcommand{\thetable}{\arabic{table}}
\renewcommand{\figurename}{Supplementary Figure}
\renewcommand{\tablename}{Supplementary Table}

\section{Dataset Composition}\label{app:dataset}

\framework builds on AssayBench~\citep{de2026assaybench}, a public benchmark containing 1,920 CRISPR screens. We adopt the same temporal train/validation/test split and retain all 1,349 training screens. From the validation and test splits, we select 20 screens each according to the following criteria:
\begin{enumerate}[noitemsep]
    \item \textbf{Genome-wide coverage}: the screen library contains between 18{,}000 and 22{,}000 genes.
    \item \textbf{Sufficient signal}: the screen contains at least 50 hits, with a hit rate $\leq 15\%$.
    \item \textbf{LLM signal floor}: the best LLM (Gemini-3-Pro) achieves AnDCG@100 $\geq 0.05$.
    \item \textbf{Phenotype stratification}: slots are allocated proportionally to the phenotype distribution in the eligible pool, with a soft cap of six screens per phenotype.
    \item \textbf{Within-phenotype diversity}: screens are selected via greedy max-min TF-IDF diversity over screen descriptions (phenotype, condition, cell line, cell type, and library methodology).
\end{enumerate}

Supplementary Table~\ref{tab:dataset_phenotype} summarizes the phenotype distribution across splits. Supplementary Table~\ref{tab:test_screens} lists all 20 test screens with their key characteristics.
Because eligibility includes a minimum LLM-signal criterion, \framework is designed to evaluate adaptive discovery on screens for which the assay description contains detectable predictive signal, rather than to represent an unbiased sample of all genome-wide CRISPR screens.

Please note that different sets of genes are used in different portions of the work. There is the full training vocabulary used to fit gene representations (23,724 genes), the fixed acquisition universe $G$ used for policy rollouts and evaluation (21,147 genes appearing in at least two screens), and each screen-specific measured library $L_s$. 

\begin{table}[h]
\centering
\small
\caption{\textbf{Phenotype distribution across \framework splits.}}
\label{tab:dataset_phenotype}
\begin{tabular}{@{}lccc@{}}
\toprule
\textbf{Phenotype category} & \textbf{Train} & \textbf{Val} & \textbf{Test} \\
\midrule
Drug / Chemical / Environmental Response & 278 & 6 & 6 \\
Fitness / Proliferation / Viability & 947 & 6 & 2 \\
Host-Pathogen / Infection Response & 38 & 6 & 5 \\
Molecular Output / Reporter / Pathway Activity & 48 & 2 & 6 \\
Trafficking / Localization / Structural Phenotypes & 38 & 0 & 1 \\
\midrule
\textbf{Total} & \textbf{1{,}349} & \textbf{20} & \textbf{20} \\
\bottomrule
\end{tabular}
\end{table}

\begin{table}[h]
\centering
\small
\caption{\textbf{Test set screens in \framework.} Each screen is a genome-wide CRISPR screen selected from the AssayBench test split. Hit rate is the fraction of genes in the library that are hits.}
\label{tab:test_screens}
\setlength{\tabcolsep}{3pt}
\begin{adjustbox}{max width=\textwidth}
\begin{tabular}{@{}llllrrrl@{}}
\toprule
\textbf{ID} & \textbf{Phenotype} & \textbf{Cell line} & \textbf{Library} & \textbf{Genes} & \textbf{Hits} & \textbf{Hit rate} & \textbf{Reference} \\
\midrule
\multicolumn{8}{@{}l}{\textit{Drug / Chemical / Environmental Response}} \\
\addlinespace
1985 & Drug response & NALM-6 & CRISPRn & 18{,}873 & 216 & 1.1\% & Krosl J (2022) \\
U\_1736\_dec & Drug response & HT-29 & CRISPRn & 19{,}008 & 1{,}189 & 6.3\% & Akinci E (2022) \\
2462 & Drug response & Capan-1 & CRISPRn & 19{,}012 & 706 & 3.7\% & Wang LM (2023) \\
1953 & Drug response & HuP-T3 & CRISPRn & 19{,}580 & 869 & 4.4\% & Hagel KR (2022) \\
2054 & Drug response & SK-N-DZ & CRISPRa & 18{,}652 & 944 & 5.1\% & Alborzinia H (2023) \\
1900 & Drug response & Primary T-cells & CRISPRn & 19{,}009 & 898 & 4.7\% & Carnevale J (2022) \\
\addlinespace
\multicolumn{8}{@{}l}{\textit{Molecular Output / Reporter / Pathway Activity}} \\
\addlinespace
U\_2423\_merged & Molecular output & Primary T-cells & CRISPRi & 18{,}811 & 225 & 1.2\% & Schmidt R (2022) \\
U\_1733\_merged & Molecular output & HeLa & CRISPRn & 18{,}385 & 169 & 0.9\% & Schraivogel D (2022) \\
U\_1987\_merged & Molecular output & HEK293T & CRISPRn & 20{,}671 & 428 & 2.1\% & Wei LH (2023) \\
U\_2422\_merged & Molecular output & Primary T-cells & CRISPRa & 18{,}800 & 571 & 3.0\% & Schmidt R (2022) \\
U\_2427\_dec & Molecular output & CD4+ T-cells & CRISPRa & 18{,}800 & 283 & 1.5\% & Schmidt R (2022) \\
2076 & Molecular output & HCT 116 & CRISPRn & 19{,}007 & 212 & 1.1\% & Rehfeld F (2023) \\
\addlinespace
\multicolumn{8}{@{}l}{\textit{Host-Pathogen / Infection Response}} \\
\addlinespace
1830 & Infection response & HEK293T/ACE2 & CRISPRn & 18{,}984 & 88 & 0.5\% & Grodzki M (2022) \\
2466 & Infection response & K-562 & CRISPRi & 18{,}734 & 88 & 0.5\% & Ngo AM (2023) \\
U\_1863\_inc & Infection response & Calu-3 & CRISPRa & 18{,}458 & 50 & 0.3\% & Rebendenne A (2022) \\
2060 & Infection response & Calu1-ACE2 & CRISPRn & 20{,}670 & 103 & 0.5\% & Ugalde AP (2022) \\
1827 & Infection response & HEK293T/ACE2 & CRISPRn & 19{,}007 & 61 & 0.3\% & Grodzki M (2022) \\
\addlinespace
\multicolumn{8}{@{}l}{\textit{Fitness / Proliferation / Viability}} \\
\addlinespace
U\_1735\_dec & Fitness & HT-29 & CRISPRn & 19{,}008 & 126 & 0.7\% & Akinci E (2022) \\
1905 & Fitness & Primary T-cells & CRISPRn & 19{,}007 & 699 & 3.7\% & Carnevale J (2022) \\
\addlinespace
\multicolumn{8}{@{}l}{\textit{Trafficking / Localization / Structural Phenotypes}} \\
\addlinespace
2404 & Trafficking & HeLa & CRISPRn & 19{,}005 & 89 & 0.5\% & Tsai PL (2022) \\
\bottomrule
\end{tabular}
\end{adjustbox}
\end{table}

\FloatBarrier
\clearpage

\section{Ablations}\label{app:ablations}

We ablate three components of \assayformer: gene embedding
initialization, the RL reward, and conditioning on the screen
description~$c_s$.

\subsection{Embedding initialization}
\label{sec:emb_init}

We compare different gene embedding initializations: BPMF (our default,
described in Section~\ref{sec:training}), GenePT~\citep{Chen2024-zm}, Singular Value Decomposition (SVD) of the training hit
matrix, Matrix Factorization (MF) and its normalized version
(MF-Sphere), and Perturb-seq derived embeddings from the K562 cell
line~\citep{littman2025gene}. As shown in
Supplementary Table~\ref{tab:ablation_results} (top), performance varies substantially
across embedding sources. After supervised fine-tuning, EF ranges from
$1.06$ (GenePT) to $3.83$ (BPMF). RL fine-tuning improves all
initializations, but the gap between BPMF and alternatives widens: SVD
is close after SFT (EF $3.72$ vs.\ $3.83$) but falls further
behind after RL ($4.40$ vs.\ $4.83$).

In Appendix~\ref{app:embedding_bio}, we further observe that downstream
performance does not correspond with how well the embeddings
recover canonical biological relationships from
STRING~\citep{szklarczyk2023string}, CORUM~\citep{giurgiu2019corum},
SIGNOR~\citep{lo2023signor}, and Reactome~\citep{milacic2024reactome},
suggesting that the structure most relevant to adaptive hit discovery
diverges from standard pathway annotations. Consistent with this, the
BPMF embeddings organize primarily by shared screen phenotype rather
than by pathway or complex membership
(Appendix~\ref{app:bpmf_structure}). The BPMF geometry is also nearly
perfectly preserved during training (Pearson $r = 0.999$;
Appendix~\ref{app:embedding_drift}), indicating that \assayformer learns
on top of this structure rather than reshaping it.

\subsection{Training strategies}
\label{sec:exp_ablations}

\begin{table}[h]
\centering
\caption{\textbf{Ablation results.} \textit{Top:} effect of gene embedding initialization on \assayformer performance after supervised fine-tuning (SFT) and RL fine-tuning. \textit{Bottom:} training strategy ablations using BPMF embeddings.}
\label{tab:ablation_results}
\setlength{\tabcolsep}{5pt}
\renewcommand{\arraystretch}{0.95}
\begin{tabular}{@{}lcc@{}}
\toprule
\textbf{Embedding initialization} & \textbf{EF} (SFT) & \textbf{EF} (RL) ($+\Delta$) \\ \midrule
Random         & $2.47$ & $2.71\;{\scriptstyle(+0.24)}$ \\
GenePT         & $1.06$ & $2.60\;{\scriptstyle(+1.54)}$ \\
K562           & $2.95$ & $3.15\;{\scriptstyle(+0.20)}$ \\
MF             & $3.01$ & $4.05\;{\scriptstyle(+1.04)}$ \\
MF-Sphere      & $3.72$ & $4.31\;{\scriptstyle(+0.59)}$ \\
SVD            & $3.72$ & $4.40\;{\scriptstyle(+0.68)}$ \\
BPMF (default) & $3.83$ & $4.83\;{\scriptstyle(+1.00)}$ \\
\midrule
\textbf{Training strategy} & \textbf{EF} & \textbf{nAUC (\%) / FH (\%)} \\ \midrule
\assayformer + GRPO                          & $4.83$ & $17.2$ / $23.2$ \\
\quad $-$ screen description & $4.52$ & $16.2$ / $22.4$ \\
\assayformer + GRPO (EF$_{\text{terminal}}$) & $4.29$ & $14.9$ / $20.5$ \\
\bottomrule
\end{tabular}
\end{table}

\paragraph{RL context-delta reward.} During RL fine-tuning, \assayformer
is trained with the context-delta reward, which explicitly rewards
improvements over a context-free policy as experimental observations
accumulate. We validate this choice against a natural alternative:
directly optimizing the terminal EF after $T=10$ acquisition rounds. As
shown in Supplementary Table~\ref{tab:ablation_results}, the terminal EF objective
leads to worse performance. Unlike the context-delta reward, it does not
isolate the benefit of conditioning on observed history, providing a
weaker learning signal for history-dependent adaptation.

\paragraph{Conditioning on the screen description $c_s$.} Because
\assayformer is primarily trained to adapt from accumulated experimental
observations, we test how much it relies on the initial screen
description. As shown in Supplementary Table~\ref{tab:ablation_results}, removing the
screen description has relatively minor effect on performance, suggesting that
initial acquisitions are driven largely by cross-screen gene hit
patterns learned from the training data rather than the text of~$c_s$.
This limited reliance on $c_s$ also helps explain the complementarity of
the handoff strategy: the LLM can exploit the screen description to
provide a biologically informed warm start, while \assayformer
subsequently adapts from experimental feedback.

\FloatBarrier
\clearpage

\section{Complete Results}\label{app:results}

\begin{table}[h!]
\centering
\caption{Full performance comparison of various baselines and ablations. Metrics evaluate enrichment factor (EF, hit rate relative to random), normalized Area Under the Curve (nAUC), fraction of hits found, mean shortfall, Vendi ratio, and batch pathway overlap versus random. EP-B/EP-S/EP-D are the effective number of Reactome level-2 pathway groups covered (186 in the vocabulary), at batch, screen and dataset scope: each gene is assigned to one of its groups at random, and the scope is subsampled to a fixed annotated-gene count (30, 200 and 6000 respectively), so the three are not directly comparable to one another.}%
\label{tab:baselines_results}
\setlength{\tabcolsep}{4pt}
\renewcommand{\arraystretch}{0.95}
\tiny
\begin{tabular}{@{}lcccccccccc@{}}
\toprule
\textbf{Method} & \textbf{EF} & \textbf{nAUC (\%)} & \textbf{Frac.\ hits (\%)} & \textbf{Shortfall (\%)} & \textbf{Ess.\ (\%)} & \textbf{Vendi (\%)} & \textbf{Path.\ Ov.} & \textbf{EP-B} & \textbf{EP-S} & \textbf{EP-D} \\ \midrule
\multicolumn{11}{@{}l}{\textit{\textbf{Base LLMs}}} \\ \addlinespace
GLM-5.1 & $4.00$ & $16.3$ & $21.0$ & $10.6$ & $28.9$ & $47.2$ & $4.72$ & $13.3$ & $30.8$ & $64.0$ \\
\quad - hit labels & $3.65$ & $15.0$ & $19.1$ & $7.3$ & $27.6$ & $47.0$ & $4.44$ & $14.4$ & $33.4$ & $64.6$ \\
GLM-5.2 & $4.09$ & $16.6$ & $21.4$ & $5.3$ & $29.1$ & $47.6$ & $4.62$ & $14.0$ & $34.0$ & $64.6$ \\
Kimi-K2.6 & $3.00$ & $14.1$ & $15.7$ & $54.2$ & $31.2$ & $60.5$ & $6.52$ & $13.2$ & $29.6$ & $64.3$ \\
\quad - hit labels & $2.72$ & $13.0$ & $14.3$ & $37.1$ & $31.4$ & $56.9$ & $4.96$ & $14.5$ & $32.9$ & $62.3$ \\
Claude Opus-4.8 & $3.86$ & $17.3$ & $20.2$ & $35.1$ & $28.3$ & $48.5$ & $6.62$ & $12.3$ & $27.4$ & $60.7$ \\
\quad - hit labels & $3.55$ & $15.8$ & $18.6$ & $28.9$ & $25.8$ & $45.4$ & $6.21$ & $12.6$ & $30.0$ & $62.6$ \\
Claude Sonnet-4.6 & $3.06$ & $13.3$ & $16.0$ & $29.7$ & $29.0$ & $51.2$ & $5.34$ & $14.1$ & $34.4$ & $65.1$ \\
Claude Haiku-4.5 & $1.39$ & $8.5$ & $7.3$ & $61.8$ & $28.9$ & $41.9$ & $6.75$ & $12.5$ & $32.7$ & $63.5$ \\
Qwen3.6-27B & $2.65$ & $10.5$ & $13.9$ & $23.4$ & $30.9$ & $41.7$ & $4.97$ & $14.3$ & $37.5$ & $68.9$ \\
\quad - hit labels & $2.32$ & $10.0$ & $12.1$ & $22.6$ & $31.2$ & $43.5$ & $5.04$ & $14.6$ & $38.1$ & $68.4$ \\
Gemini-3.1-pro & $4.71$ & $20.6$ & $24.6$ & $3.6$ & $30.6$ & $47.4$ & $4.39$ & $14.6$ & $34.6$ & $66.2$ \\
\quad - hit labels & $4.38$ & $18.5$ & $22.9$ & $3.1$ & $29.5$ & $47.1$ & $4.53$ & $14.3$ & $34.3$ & $63.5$ \\
GPT-5.5 & $4.43$ & $18.7$ & $23.1$ & $5.1$ & $27.2$ & $46.6$ & $5.02$ & $12.2$ & $28.2$ & $60.2$ \\
GPT-5.6 Sol & $4.81$ & $20.5$ & $25.2$ & $5.0$ & $29.6$ & $49.7$ & $4.54$ & $13.2$ & $29.5$ & $61.8$ \\
\midrule
\multicolumn{11}{@{}l}{\textit{\textbf{AssayLLM (Ours)}}} \\ \addlinespace
Qwen3.6-27B (base) & $2.65$ & $10.5$ & $13.9$ & $23.4$ & $30.9$ & $41.7$ & $4.97$ & $14.3$ & $37.5$ & $68.9$ \\
\quad + SFT (GLM-5.1 traces) & $3.56$ & $14.7$ & $18.7$ & $19.0$ & $35.9$ & $42.2$ & $5.36$ & $13.6$ & $34.3$ & $64.1$ \\
\quad + SFT + GRPO (= AssayLLM) & $3.69$ & $15.5$ & $19.3$ & $16.7$ & $37.8$ & $41.4$ & $5.52$ & $13.7$ & $34.7$ & $64.8$ \\
\midrule
\multicolumn{11}{@{}l}{\textit{\textbf{Adaptive Experimental Design Methods}}} \\ \addlinespace
Random & $1.04$ & $3.4$ & $4.9$ & $10.0$ & $18.6$ & $58.5$ & $1.10$ & $21.7$ & $56.9$ & $82.6$ \\
Prior hit baseline & $2.87$ & $11.5$ & $14.9$ & $1.5$ & $99.1$ & $56.5$ & $3.22$ & $17.8$ & $38.4$ & $48.4$ \\
Screen-kNN & $3.40$ & $12.5$ & $17.3$ & $3.7$ & $77.1$ & $57.6$ & $2.91$ & $18.9$ & $42.8$ & $54.8$ \\
RF (greedy) & $1.99$ & $6.4$ & $9.9$ & $3.7$ & $30.5$ & $50.1$ & $3.24$ & $18.4$ & $49.2$ & $77.3$ \\
RF + UCB & $1.32$ & $3.8$ & $6.2$ & $10.9$ & $23.9$ & $47.1$ & $2.42$ & $18.1$ & $47.8$ & $74.0$ \\
BPMF~\citep{Salakhutdinov2008-ad} & $4.49$ & $12.3$ & $19.7$ & $18.0$ & $36.1$ & $53.4$ & $2.05$ & $20.1$ & $50.7$ & $72.5$ \\
Transformer + DAgger~\citep{Ross2011-wr} & $2.80$ & $10.3$ & $14.4$ & $2.5$ & $90.0$ & $59.8$ & $2.52$ & $19.0$ & $41.9$ & $53.9$ \\
MAML (w/ BPMF embs.)~\citep{finn2017model,Salakhutdinov2008-ad} & $3.77$ & $14.3$ & $18.6$ & $6.4$ & $74.9$ & $58.2$ & $2.54$ & $19.4$ & $44.4$ & $57.8$ \\
BioBO~\citep{li2025biobo} & $2.59$ & $5.9$ & $12.2$ & $10.3$ & $38.4$ & $64.3$ & $1.68$ & $19.9$ & $50.6$ & $72.6$ \\
Probability-of-hit~\citep{Rubbi2026-oo} & $2.41$ & $7.3$ & $11.8$ & $9.0$ & $35.2$ & $56.6$ & $2.11$ & $20.3$ & $53.4$ & $79.0$ \\
\midrule
\multicolumn{11}{@{}l}{\textit{\textbf{Agent Harnesses}}} \\ \addlinespace
Haiku-4.5 Agent & $3.20$ & $14.3$ & $16.8$ & $20.6$ & $49.8$ & $61.7$ & $2.94$ & $18.6$ & $47.3$ & $68.3$ \\
LLMNN~\citep{Gupta2025-qb} & $2.39$ & $9.4$ & $12.5$ & $0.1$ & $22.6$ & $64.9$ & $1.44$ & $20.3$ & $52.1$ & $78.6$ \\
ICBR-EF~\citep{Wainrib2026-rs} & $2.76$ & $9.3$ & $14.5$ & $0.9$ & $32.2$ & $61.5$ & $2.32$ & $18.9$ & $44.1$ & $68.5$ \\
\midrule
\multicolumn{11}{@{}l}{\textit{\textbf{AssayFormer (Ours)}}} \\ \addlinespace
AssayFormer (random embs.) & $2.47$ & $9.2$ & $12.3$ & $5.4$ & $86.1$ & $56.9$ & $2.68$ & $18.5$ & $40.0$ & $51.5$ \\
\quad + BPMF Embeddings & $3.83$ & $13.2$ & $18.3$ & $10.2$ & $37.5$ & $52.5$ & $2.54$ & $18.3$ & $42.7$ & $56.9$ \\
\quad + GRPO (= AssayFormer) & $4.83$ & $17.2$ & $23.2$ & $10.0$ & $40.7$ & $55.9$ & $2.38$ & $19.3$ & $46.7$ & $65.0$ \\
\midrule
\multicolumn{11}{@{}l}{\textit{\textbf{AssayLoop (Ours)}}} \\ \addlinespace
GLM-5.1 $\rightarrow$ AssayFormer & $5.27$ & $19.8$ & $25.5$ & $8.5$ & $31.6$ & $55.9$ & $2.88$ & $18.8$ & $47.5$ & $71.0$ \\
Gemini-3.1-Pro $\rightarrow$ AssayFormer & $5.67$ & $21.7$ & $27.7$ & $7.6$ & $30.1$ & $55.6$ & $3.06$ & $18.4$ & $46.6$ & $70.4$ \\
GPT-5.5 $\rightarrow$ AssayFormer & $5.51$ & $20.6$ & $26.8$ & $7.9$ & $29.5$ & $54.7$ & $3.28$ & $17.8$ & $45.4$ & $69.6$ \\
GPT-5.6 Sol $\rightarrow$ AssayFormer & $5.67$ & $21.7$ & $27.6$ & $7.8$ & $29.4$ & $55.8$ & $3.06$ & $18.1$ & $45.9$ & $69.8$ \\
AssayLLM $\rightarrow$ AssayFormer & $5.05$ & $18.6$ & $24.7$ & $7.8$ & $34.8$ & $55.5$ & $2.97$ & $18.8$ & $47.5$ & $70.8$ \\
Handoff-trained AssayLLM $\rightarrow$AssayFormer & $5.23$ & $18.7$ & $25.1$ & $9.1$ & $39.2$ & $53.7$ & $2.91$ & $18.9$ & $47.0$ & $68.3$ \\
\midrule
\multicolumn{11}{@{}l}{\textit{\textbf{Ablations}}} \\ \addlinespace
AssayFormer (Random) & $2.47$ & $9.2$ & $12.3$ & $5.4$ & $86.1$ & $56.9$ & $2.68$ & $18.5$ & $40.0$ & $51.5$ \\
AssayFormer (BPMF) & $3.83$ & $13.2$ & $18.3$ & $10.2$ & $37.5$ & $52.5$ & $2.54$ & $18.3$ & $42.7$ & $56.9$ \\
\quad - screen desc. & $3.82$ & $13.1$ & $19.1$ & $5.1$ & $62.6$ & $56.7$ & $2.68$ & $18.8$ & $43.2$ & $56.5$ \\
AssayFormer (SVD) & $3.72$ & $10.3$ & $11.1$ & $46.5$ & $44.1$ & $24.7$ & $4.62$ & $19.0$ & $45.5$ & $61.1$ \\
AssayFormer (MF) & $3.01$ & $8.9$ & $10.3$ & $36.3$ & $26.1$ & $34.6$ & $2.94$ & $20.2$ & $50.0$ & $69.2$ \\
AssayFormer (MF-Sphere) & $3.72$ & $11.1$ & $13.0$ & $35.9$ & $33.4$ & $30.8$ & $2.76$ & $20.0$ & $48.9$ & $67.5$ \\
AssayFormer (GenePT-PCA) & $1.06$ & $4.2$ & $4.7$ & $15.4$ & $22.7$ & $33.9$ & $5.55$ & $18.5$ & $44.7$ & $54.4$ \\
AssayFormer (K562-PCA) & $2.95$ & $10.2$ & $14.0$ & $10.6$ & $55.3$ & $58.7$ & $2.16$ & $19.8$ & $48.4$ & $64.5$ \\
\addlinespace
Random\hfill + GRPO & $2.71$ & $9.8$ & $12.7$ & $10.9$ & $71.0$ & $54.9$ & $1.99$ & $19.9$ & $46.3$ & $60.8$ \\
BPMF\hfill + GRPO & $4.83$ & $17.2$ & $23.2$ & $10.0$ & $40.7$ & $55.9$ & $2.38$ & $19.3$ & $46.7$ & $65.0$ \\
BPMF (no desc.)\hfill + GRPO & $4.52$ & $16.2$ & $22.4$ & $5.9$ & $42.5$ & $57.7$ & $2.55$ & $18.7$ & $45.7$ & $63.1$ \\
BPMF\hfill + GRPO (EF$_{\text{terminal}}$) & $4.29$ & $14.9$ & $20.5$ & $10.7$ & $48.4$ & $54.2$ & $2.35$ & $19.0$ & $45.0$ & $62.7$ \\
SVD\hfill + GRPO & $4.40$ & $16.0$ & $22.9$ & $1.8$ & $36.5$ & $63.4$ & $2.09$ & $19.4$ & $46.6$ & $64.6$ \\
MF\hfill + GRPO & $4.05$ & $13.8$ & $18.9$ & $15.8$ & $46.5$ & $45.5$ & $3.12$ & $17.3$ & $38.7$ & $52.8$ \\
MF-Sphere\hfill + GRPO & $4.31$ & $14.6$ & $20.4$ & $11.7$ & $39.0$ & $52.4$ & $2.27$ & $18.8$ & $45.0$ & $63.3$ \\
GenePT-PCA\hfill + GRPO & $2.60$ & $9.3$ & $12.2$ & $10.4$ & $73.9$ & $38.1$ & $11.46$ & $16.1$ & $35.4$ & $45.0$ \\
K562-PCA\hfill + GRPO & $3.15$ & $11.6$ & $15.7$ & $6.5$ & $79.1$ & $55.8$ & $3.79$ & $18.1$ & $41.7$ & $54.2$ \\
\bottomrule
\end{tabular}
\end{table}

\FloatBarrier
\newpage

\section{Implementation details}

\subsection{Gene embeddings initialization}
\label{app:bpmf}

Let $Y_{ij}\in\{0,1\}$ indicate
whether gene $j$ scored as a hit in screen $i$, observed on the set $\mathcal{O}$
of screen--gene pairs actually assayed ($N=1{,}349$ training screens,
$M=23{,}724$ genes (the total number in our training data), $64\%$ of pairs observed, overall hit rate $7.3\%$). Each
entry is generated by a latent utility
\begin{equation}
  Z_{ij} = U_i^{\top} V_{g_j} + \varepsilon_{ij},\quad
  \varepsilon_{ij}\sim\mathcal{N}(0,1),
  \qquad
  Y_{ij} = \mathbb{I}\!\left[Z_{ij} > 0\right],
\end{equation}
with isotropic Gaussian priors $U_i\sim\mathcal{N}(0,\sigma_u^2 I_{d_g})$ on the
screen factors and $V_{g_j}\sim\mathcal{N}(0,\sigma_v^2 I_{d_g})$ on the gene
factors ($\sigma_u=\sigma_v=1$). Conditioning on $Z$ renders both factor updates
conjugate, so we alternate three Gibbs steps: $Z \mid Y,U,V$ drawn from
unit-variance normals truncated at zero by the sign of $Y$
\citep{albert1993bayesian}, followed by
\begin{equation}
  U_i \mid Z,V \;\sim\; \mathcal{N}\!\left(\Lambda_i^{-1} \!\!\!
    \sum_{j:(i,j)\in\mathcal{O}} \!\!\! V_{g_j} Z_{ij},\;\; \Lambda_i^{-1}\right),
  \qquad
  \Lambda_i = \!\!\! \sum_{j:(i,j)\in\mathcal{O}} \!\!\! V_{g_j} V_{g_j}^{\top}
    + \sigma_u^{-2} I_{d_g},
\end{equation}
and the symmetric update for $V_{g_j} \mid Z,U$. Unobserved pairs drop out of
these sums, so the fit is a matrix \emph{completion} rather than an imputation of
zeros. We run $2{,}000$ iterations, discard the first $1{,}000$ as burn-in, thin
by $2$, and set $V_{g_j}\in\mathbb{R}^{d_g}$ with $d_g = K = 10$ to the posterior
mean of the retained samples. Genes absent from the factorization receive
$V_{g}=\mathbf{0}$ and a static bias $\mathrm{logit}(\hat{p}_g)$ at their marginal
hit frequency; genes covered by the fit carry no per-gene bias at initialization, so their score is
the pure bilinear form $\hat{u}^{\top} V_{g}$ and all ranking signal must come
from the screen latent $\hat{u}$ that \assayformer infers in context.

\subsection{GRPO}
Advantages are calculated as the
remaining returns $G_t^{(i)} = \sum_{s \geq t} r_s^{(i)}$ and normalized across the
group at each step, $A_t^{(i)} = (G_t^{(i)} - \mu_t)/\sigma_t$. 
The policy is then updated by the REINFORCE gradient of the Plackett–Luce log-probabilities weighted by $A_t^{(i)}$, with an entropy bonus (coef $0.01$). 
Notably, the learned policy during RL acquires from and is evaluated on the full gene universe (including genes without hit labels in the given screen), so the policy learns to rank the full candidate space, not just each screen's library.

\subsection{Training settings} In the SFT phase, \assayformer was trained for 40 epochs with a batch size of 32 screens, AdamW, lr=$3e-4$, and weight decay $1e-2$. Gene embeddings were initialized by BPMF ($d_g$=10) and left trainable (with no weight decay). The best validation epoch was selected for post-training. It was post-trained with GRPO across 4 B200 GPUs for 100 epochs with group size 8, $\gamma = 1.0$, temperature 1.0. There were ($\text{floor}(1349/4)=337$) steps per epoch. Rollouts were conducted using the universe of genes appearing in two or more screens (21,147 genes, same as evaluation). The context delta reward was used, with the frozen policy $\pi_0$ updated to the current policy every 25 epochs. Training used AdamW with kl\_coef 0.0, ent\_coef 0.01, lr 1e-5, weight decay 0.0, grad-norm clipping to 1.0. This resulted in 1348 ($\lfloor$ training set size $ / 4 \rfloor \times 4$) * 8 (group size) * 100 (epochs) = 1,078,400 traces during training. This is equivalent to 10,784,000 rounds of active learning and 1,078,400,000 genes acquired. We selected a final model checkpoint used in test-set evaluation was selected using validation handoff performance. 

\subsection{\assayllm{}}\label{app:assayllm}
\paragraph{Backbone.} Qwen3.6-27B, a dense $27$B open-weight model with hybrid
linear/full attention. Training uses SDPA attention throughout (FlashAttention-2
is numerically unstable for this architecture on our accelerators) and therefore
no sequence packing. All stages use \texttt{bf16} parameters with DeepSpeed
ZeRO-3; ZeRO-2 exceeds host memory for a model of this size.

\subsubsection{SFT}
\label{app:sft}

\paragraph{Trajectory construction.} Teacher runs are $T$-round campaigns, one
row per (run, screen, round). We reassemble them into single multi-turn
conversations: a shared system prompt describing the goal, screen context,
objective, hit definition, ranking criteria, and required output format; a
$t{=}1$ user turn carrying the screen description $c_s$; then, per round, an
assistant turn holding reasoning traces, an answer $B_t$ as a ranked comma-separated list of $b{=}100$ genes,
and a user turn holding incremental feedback. Feedback lists the labels newly
added to $h_t$, the running cumulative counts, and a non-repetition instruction.
Rows whose assistant body is not a clean gene list are dropped (roughly $3\%$).
Only assistant tokens are unmasked in the loss.

\paragraph{Optimization.} Full-parameter fine-tuning across $16$ GPUs with ZeRO-3, gradient checkpointing, and SDPA attention;
$3$ epochs, global batch size $32$, learning rate
$1{\times}10^{-6}$ with $3\%$ warmup.

\subsubsection{GRPO}
\label{app:grpo}
\paragraph{Rollout.} One prompt is one screen. The policy generates
$B_1,\ldots,B_k$ with reasoning enabled; after each round the environment parses
the gene list, reveals $y_g$ for picks in $\mathcal{G}_s$, and appends the
incremental feedback turn. Group size is $16$ samples per prompt, sampled at
temperature $1.0$ and top-$p\,0.95$; higher diversity is useful here because
group-relative advantages need within-group spread.

\paragraph{Optimization.} We then optimize the policy with Group Relative Policy Optimization (GRPO). For each prompt $x$ we sample a group of $G$ responses $\{y_i\}_{i=1}^{G}$ from the current policy
and compute the group-normalized advantage
\begin{equation}
\hat{A}_i \;=\; \frac{r_i - \mathrm{mean}(\{r_j\}_{j=1}^{G})}
                     {\mathrm{std}(\{r_j\}_{j=1}^{G})},
\end{equation}
and maximize the clipped surrogate objective
\begin{equation}
\mathcal{J}(\theta) =
\mathbb{E}\!\left[\frac{1}{G}\sum_{i=1}^{G}
\min\!\Big(
\rho_i \hat{A}_i,\;
\mathrm{clip}(\rho_i, 1-\epsilon, 1+\epsilon)\hat{A}_i
\Big)\right]
- \beta\, \mathbb{D}_{\mathrm{KL}}\!\left(\pi_\theta \,\|\, \pi_{\mathrm{ref}}\right),
\end{equation}
where $\rho_i = \pi_\theta(y_i\mid x)/\pi_{\theta_{\text{old}}}(y_i\mid x)$.
Group-normalized advantages, global batch size $48$, actor learning rate $1{\times}10^{-6}$ with $3\%$
warmup, ZeRO-3 / \texttt{bf16}, non-reentrant gradient checkpointing, SDPA
attention.

\subsection{Screen-kNN implementation}\label{app:screen_knn}

\screenknn predicts gene hits by finding the most similar historical screens
in the training set and aggregating their outcomes. Let $\mathcal{S}$ denote
the set of training screens. Following the notation of the main text, each
training screen $s \in \mathcal{S}$ has a gene library $L_s \subseteq
\mathcal{G}$ and binary hit labels $y_g^s \in \{0, 1\}$ for $g \in L_s$.

\paragraph{Prior.} We compute a global prior for each gene as its empirical
hit rate across all training screens in which it appears:
\begin{equation}
    p_0(g) = \frac{\sum_{s \in \mathcal{S}} y_g^s \, \mathbb{I}[g \in L_s]}
                   {\sum_{s \in \mathcal{S}} \mathbb{I}[g \in L_s]}.
\end{equation}

\paragraph{Screen similarity.} At round $t$, the policy conditions on the
history $h_{t-1}$, whose gene set $G_{t-1}$ carries the labels $\{y_g\}_{g \in
G_{t-1}}$ observed so far in the current screen. For each training screen $s$,
we compute the normalized Hamming distance over the genes shared with that
history:
\begin{equation}
    \delta(s) = \frac{1}{|G_{t-1} \cap L_s|}
                \sum_{g \in G_{t-1} \cap L_s} (y_g - y_g^s)^2 .
\end{equation}
If $G_{t-1} \cap L_s = \emptyset$, we set $\delta(s) = 1$. Distances are
converted to weights via an exponential kernel with temperature $\tau$:
\begin{equation}
    w_s = \exp\!\left(-\frac{\delta(s)}{\tau}\right).
\end{equation}

\paragraph{Prediction.} The screen evidence for gene $g$ is the weighted hit
rate across training screens whose library contains $g$:
\begin{equation}
    p_{\text{screen}}(g) = \frac{\sum_{s\,:\,g \in L_s} w_s \, y_g^s}
                                 {\sum_{s\,:\,g \in L_s} w_s}.
\end{equation}
The acquisition score is a Bayesian-style mixture of the prior and the screen
evidence,
\begin{equation}
    a_{\textsc{knn}}(g \mid h_{t-1}) = (1 - \lambda_t) \, p_0(g)
        + \lambda_t \, p_{\text{screen}}(g),
\end{equation}
where $\lambda_t = |G_{t-1}| / (|G_{t-1}| + \kappa)$ interpolates between the
prior (when few genes have been revealed) and the screen evidence (as more
observations accumulate). We use $\tau = 0.1$ and $\kappa = 100$, and form the
batch $B_t$ from the $b$ untested genes with the highest score.

\newpage

\section{Example Inputs}

\subsection{LLM Prompt Examples}\label{app:prompt_template}

The following is an example active-learning prompt AFTER round 1 for the following screen: 
\begin{itemize}
\item Screen: 2054 
\item Cell line: SK-N-DZ (Neuroblastoma)
\item CRISPRa
\item RSL3 ferroptosis-resistance screen
\item TEST set. 
\end{itemize}

\begin{tcolorbox}
\textbf{System Prompt}
\begin{lstlisting}[basicstyle=\scriptsize\ttfamily, breaklines=true, breakindent=0pt, columns=fullflexible, literate={—}{---}1]
You are a computational biologist. Given a CRISPR screen description, produce a ranked list of 100 gene symbols (HGNC nomenclature for Homo sapiens) that you believe are MOST LIKELY to be hits in this screen, ordered strongest first. Use only your knowledge of biology — no chain-of-thought in the output. Return ONLY a comma-separated list of 100 symbols, no prose, no numbering.
\end{lstlisting}
\end{tcolorbox}

\vspace*{-15mm}
\begin{tcolorbox}
\textbf{User Prompt}
\begin{lstlisting}[basicstyle=\scriptsize\ttfamily, breaklines=true, breakindent=0pt, columns=fullflexible]
## Goal

You are tasked with ranking genes from a genetic perturbation screen. Based on the experimental context and hit criteria provided below, provide a list of exactly 100 genes that are hits in this screen, ranked from strongest to weakest according to the criteria defined below.

## Experimental Context

This screen was performed in SK-N-DZ cells, a Neuroblastoma Cell Line. Researchers used a CRISPRa library (Activation) to systematically perturb gene function. The experiment followed a Drug Exposure design and was conducted over 14 Days under 1S,3R-RSL3 treatment (300.0 nM).

## Screen Objective

The primary objective of this screen was to identify a set of hit genes, each of which increases resistance to the ferroptosis-inducing drug RSL3, as measured by increased cell viability after exposure.

## Hit Definition

A gene is classified as a "hit" if its Activation significantly increases resistance to the ferroptosis-inducing drug RSL3, as measured by increased cell viability after exposure. The statistical criterion for significance is: FDR < 0.05.

## Ranking Criteria

Genes with low FDR are ranked most highly.

## Additional Context
Screen notes: A genome-wide CRISPR-activation (CRISPRa) screen was performed to identify novel regulators of ferroptosis. This CRISPRa_RSL3 screen involved positive selection for viability after exposing the cells to the ferroptosis inducing agent RSL-3 which inhibits GPX4

## Required Output Format

Provide your response as an ordered list of exactly 100 HGNC gene symbols, using the ranking criteria above.
That is, top genes should have low FDR.

Format: 
GENE1, GENE2, GENE3, ..., GENE100

## Active-Learning History

You have already validated 99 gene(s) in this screen, shown below grouped by round in acquisition order. Do NOT re-suggest any of them.

### Round 1 (99 genes, hit rate 12/99 = 12.1%
  Hits (12): GPX4, AIFM2, GCH1, MBOAT2, GCLM, LRP8, SLC7A11, DGAT1, CHMP4A, NQO1, ME1, MUC1
  Non-hits (87): DHODH, MBOAT1, SCD, ACSL3, FTH1, FTL, NFE2L2, AKR1C1, AKR1C2, AKR1C3, PROM2, SLC40A1, COQ2, COQ6, COQ8A, COQ9, PDSS1, PDSS2, PTS, SPR, DHFR, QDPR, GSS, GCLC, TXNRD1, TXN, SQSTM1, HMOX1, MT1G, HSPA5, HSPB1, ALDH3A1, SECISBP2, SEPSECS, EEFSEC, SEPHS2, SLC3A2, CAV1, ABCB1, ABCG2, ABCC1, PLIN2, PLIN3, PLIN4, DGAT2, SREBF1, FASN, CHMP5, CHMP6, VPS4A, VPS4B, CHMP1A, CHMP1B, CHMP2A, CHMP2B, CHMP3, CHMP4B, CHMP7, IST1, PRDX1, PRDX6, SRXN1, GSR, IDH1, PGD, G6PD, TKT, TALDO1, ATF4, ASNS, PCK2, OTUB1, CD44, SOX9, HIF1A, EIF2S1, EIF2AK3, EIF2AK4, PCBP1, PCBP2, COQ3, COQ4, COQ5, COQ7, MVD, FDPS, HMGCR

  Cumulative: 12/99 hits (12.1%


=== Final instruction ===
Return ONLY a comma-separated list of 100 gene symbols (no prose, no numbers, no markdown). All gene symbols must use the HGNC convention for Homo sapiens (e.g. 'TP53' for HGNC, 'Trp53' for MGI). Do not return symbols from another organism.
\end{lstlisting}
\end{tcolorbox}

\begin{tcolorbox}
\textbf{User Prompt History without Labels}
\begin{lstlisting}[basicstyle=\scriptsize\ttfamily, breaklines=true, breakindent=0pt, columns=fullflexible]
[...]

## Active-Learning History

You have already sampled 99 gene(s) in this screen, shown below grouped by round in acquisition order. Hit/non-hit labels are NOT shown. Do NOT re-suggest any of them.

### Round 1 (99 genes sampled)
  Sampled: GPX4, AIFM2, GCH1, DHODH, MBOAT1, MBOAT2, SCD, ACSL3, FTH1, FTL, NFE2L2, AKR1C1, AKR1C2, AKR1C3, PROM2, SLC40A1, COQ2, COQ6, COQ8A, COQ9, PDSS1, PDSS2, PTS, SPR, DHFR, QDPR, GSS, GCLC, GCLM, TXNRD1, TXN, SQSTM1, HMOX1, MT1G, HSPA5, HSPB1, ALDH3A1, SECISBP2, SEPSECS, EEFSEC, SEPHS2, LRP8, SLC3A2, SLC7A11, CAV1, ABCB1, ABCG2, ABCC1, PLIN2, PLIN3, PLIN4, DGAT1, DGAT2, SREBF1, FASN, CHMP5, CHMP6, VPS4A, VPS4B, CHMP1A, CHMP1B, CHMP2A, CHMP2B, CHMP3, CHMP4A, CHMP4B, CHMP7, IST1, NQO1, PRDX1, PRDX6, SRXN1, GSR, ME1, IDH1, PGD, G6PD, TKT, TALDO1, ATF4, ASNS, PCK2, OTUB1, CD44, MUC1, SOX9, HIF1A, EIF2S1, EIF2AK3, EIF2AK4, PCBP1, PCBP2, COQ3, COQ4, COQ5, COQ7, MVD, FDPS, HMGCR

[...]
\end{lstlisting}
\end{tcolorbox}

\subsection{Qualitative analysis of \assayloop on two biologically distinct screens}

\begin{figure}[htbp]
    \centering
    \includegraphics[width=1\textwidth]{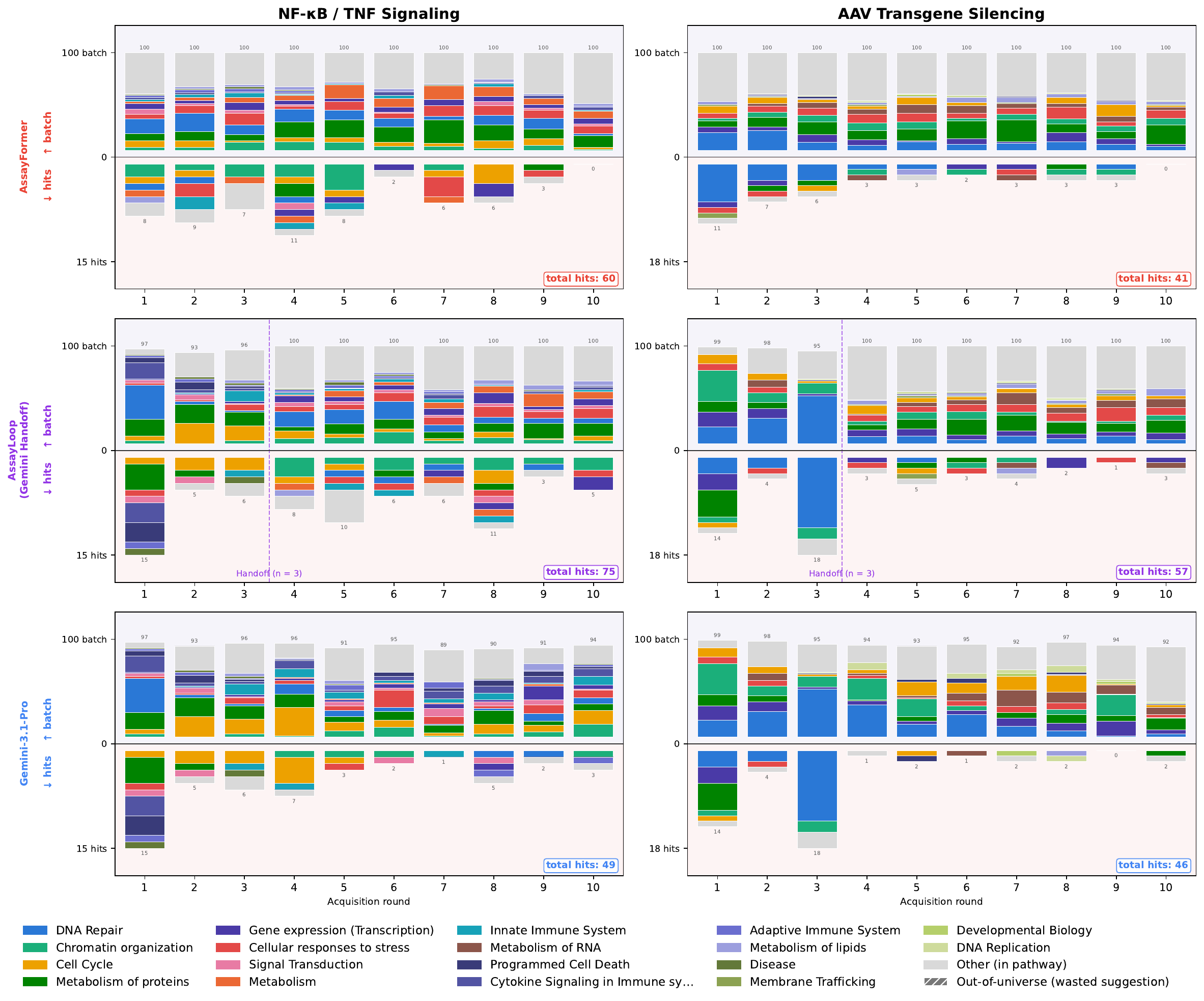}
    \caption{\textbf{Qualitative Analysis of \assayloop on two biologically distinct screens.} We show the composition of genes of suggested acquisition rounds across three models: \assayformer (top row), Gemini-3.1-Pro (bottom row), and the \assayloop handoff from Gemini (middle row). On the left side of the figure, we have a $\text{NF-}\kappa\text{B}$ / TNF signaling screen. On the right side, we have a AAV transgene silencing screen. For each subplot (representing a screen-model pair), we have two histograms. The upper histogram shows the proposed acquisition batch at each acquisition round. The lower histogram shows the composition of hits from that batch. For example, in the AAV screen, \assayformer proposes 100 genes for the first batch, with DNA repair being the largest category. The bars are colored according to gene pathways from Reactome.  }
    \label{fig:QA_composition}
\end{figure}

\FloatBarrier
\newpage

\section{Gene Embedding Analysis}\label{app:embeddings}

This appendix provides additional analysis of the gene embeddings used
by \assayformer, complementing the ablation in
Section~\ref{sec:emb_init}.

\subsection{Embedding drift during training}\label{app:embedding_drift}

A natural question is whether \assayformer reshapes the gene embedding
geometry during training or preserves the structure imposed by the
initialization. We examine this for all ablated embedding types by
measuring two complementary quantities: pairwise cosine similarity
preservation and gene-neighborhood stability.

\begin{figure}[htbp]
    \centering
    \includegraphics[width=0.68\textwidth]{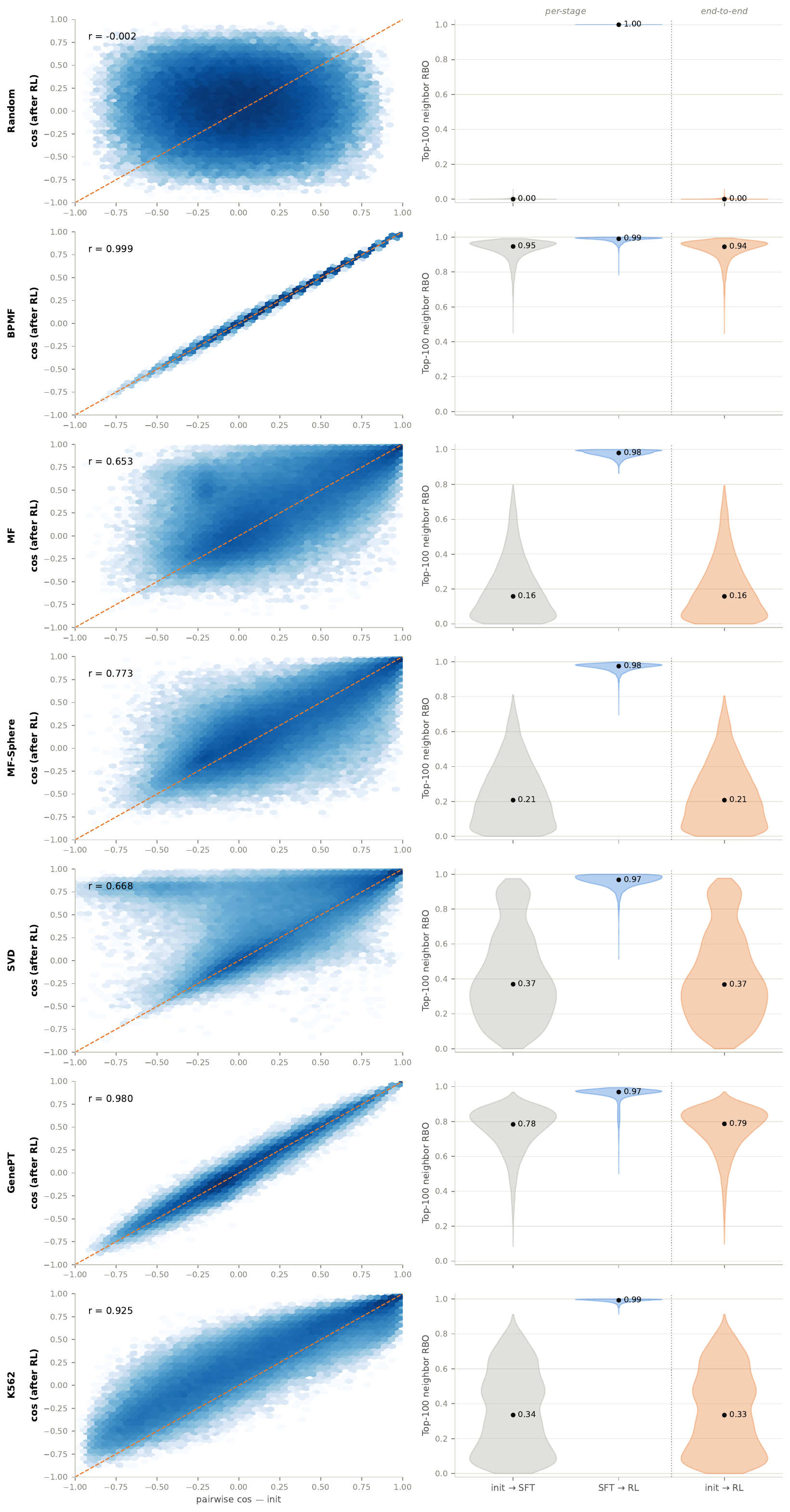}
    \caption{\small \textbf{Embedding drift during training for all ablated embedding types.} The effect of training on embedding geometry differs dramatically by embedding type. BPMF embeddings show near-perfect preservation of gene neighborhoods, while other initializations undergo substantial reorganization during supervised fine-tuning.}
    \label{fig:embedding_drift_source}
\end{figure}

\paragraph{Pairwise cosine similarity.} For each embedding type, we
compute the pairwise cosine similarity between 300,000 randomly sampled
gene pairs before and after training. As reported in the main text
(Supplementary Fig.~\ref{fig:embedding_drift_source}), BPMF embeddings are almost perfectly
preserved (Pearson $r = 0.999$). This indicates that the geometry
provided by BPMF is already well-suited for the acquisition task and
that \assayformer learns its policy on top of this fixed structure
rather than reorganizing it.

\paragraph{Gene-neighborhood stability.} We further quantify structural
changes using rank-biased overlap
(RBO)~\citep{webber2010similarity} between the 100 nearest neighbors of
each gene before and after each training stage. For BPMF, the median RBO
between initialization and the final RL model exceeds $0.95$, confirming
that gene neighborhoods are largely unchanged.

Supplementary Fig.~\ref{fig:embedding_drift_source} shows the degree of embedding
drift across all ablated embedding types. The drift varies
dramatically across sources. Random embeddings undergo near-complete
reorganization during supervised fine-tuning, as expected. Other
matrix-factorization variants (SVD, MF, MF-Sphere) also show
substantial neighborhood changes during supervised fine-tuning, though
less than random initialization. Interestingly, GenePT embeddings show
relatively little drift despite their poor downstream performance,
suggesting that embedding stability alone does not explain the advantage
of BPMF. In all cases, the RL fine-tuning stage produces much smaller
changes than supervised fine-tuning.

Together, these results suggest that the effectiveness of BPMF stems not
merely from stability but from providing a geometry that is both stable
\emph{and} well-aligned with the adaptive hit-discovery objective.
Other embeddings are either reshaped during training (random, SVD, MF)
without converging to an equally effective geometry, or are stable but
encode structure less relevant to the task (GenePT).

\subsection{Embedding initialization and biological plausibility}\label{app:embedding_bio}

\begin{figure}[htbp]
    \centering
    \includegraphics[width=0.8\textwidth]{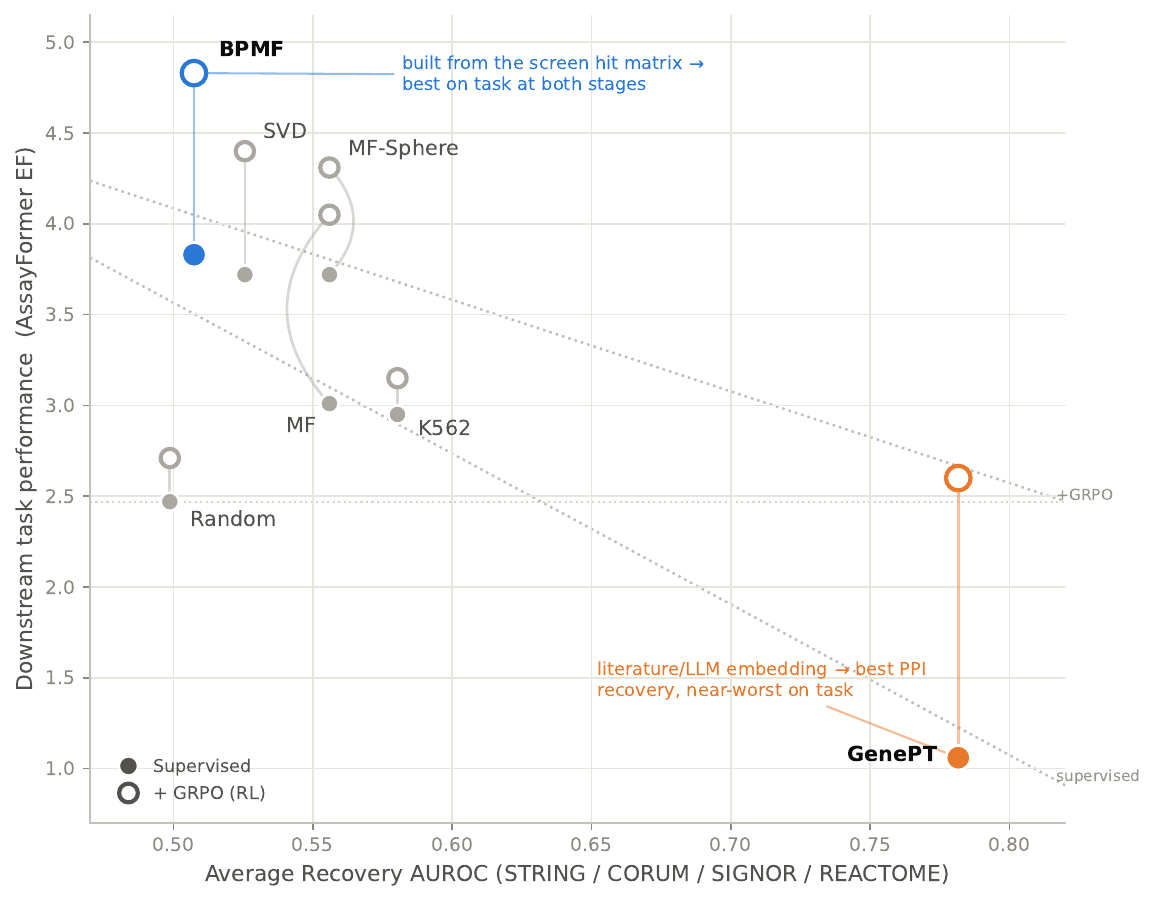}
    \caption{\textbf{Recovering textbook biology does not predict task usefulness.} We take initial embeddings and measure how much they recover gene-gene relationships from four well known databases: STRING, CORUM, SIGNOR, and REACTOME. We compute gene-gene similarity scores with cosine similarity. Ground truth labels are assigned to be 1 if the relationship exists in the database, otherwise 0. AUROC is calculated using these scores and then averaged across the four databases. Interestingly, we find that embeddings which capture these databases most closely (e.g., GenePT, K562 embeddings) are more difficult to train \assayformer on, whereas methods which initialize directly from historical screen data (matrix factorization approaches) are most useful. }
    \label{fig:embedding_story}
\end{figure}

As discussed in Section~\ref{sec:emb_init}, we observe a negative
association between how well an embedding source recovers canonical
biological relationships and its downstream performance as an
\assayformer initialization. Supplementary Fig.~\ref{fig:embedding_story} visualizes
this relationship. For each embedding type, we compute pairwise
cosine similarity between genes and measure the AUROC for recovering
known gene-gene relationships from STRING~\citep{szklarczyk2023string},
CORUM~\citep{giurgiu2019corum}, SIGNOR~\citep{lo2023signor}, and
Reactome~\citep{milacic2024reactome}, averaged across the four
databases. GenePT and K562 embeddings, which capture these known
relationships best, provide the weakest initialization for \assayformer.
Conversely, matrix-factorization approaches that learn directly from
historical screen data perform best, despite encoding less canonical
biology. This suggests that the gene-gene structure most relevant to
adaptive hit discovery is not well captured by established pathway and
interaction databases.

\subsection{Structure of the BPMF embedding space}\label{app:bpmf_structure}

\begin{figure}[htbp]
    \centering
    \includegraphics[width=.95\textwidth]{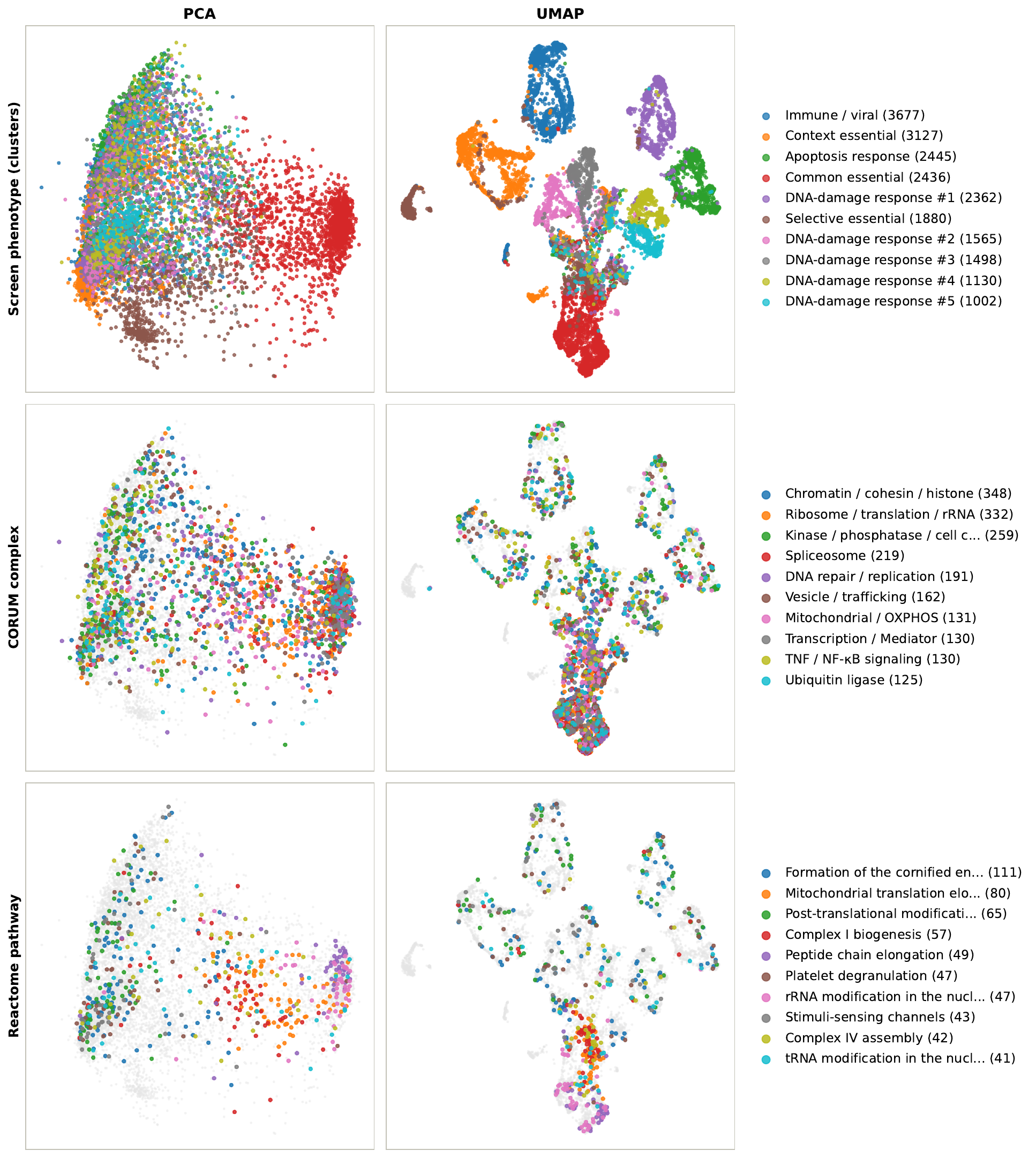}
    \caption{\textbf{BPMF embeddings organize by screen phenotype, rather than physical complex or pathway.} Embeddings are $K=10$ dimensions and are visualized with PCA (left) and cosine UMAP (right). %
    \textbf{Top: HDBSCAN}. Genes are clustered in the original space with HDBSCAN \citep{McInnes2017} with minimum cluster size 150, each labeled by its dominant phenotype association. Note that several clusters are associated with DNA-damage response without further disambiguation. \textbf{Middle: CORUM Complexes}. The embeddings are colored by CORUM complexes grouped into general families (top 10 groups, grey = other). This coloring shows minimal grouping across the space. \textbf{Bottom: Reactome pathways.} The largest 10 Reactome pathways are used to color the embeddings, similarly revealing a lack of distinct spatial clustering except in the common essential gene area. 
}
    \label{fig:bpmf_embedding}
\end{figure}

To better understand why BPMF embeddings are effective, we visualize and
cluster the $d_g=10$-dimensional embedding space. Supplementary Fig.~\ref{fig:bpmf_embedding}
shows PCA and cosine-UMAP projections of the BPMF gene embeddings,
colored by three different annotation schemes.

\paragraph{HDBSCAN clustering.} We cluster genes in the original
embedding space using HDBSCAN~\citep{McInnes2017} with a minimum cluster
size of 150. The resulting clusters align primarily with screen
phenotype categories rather than molecular pathway or complex
membership. We identify clusters corresponding to common essential,
context-essential, and selectively essential genes, as well as clusters
associated with immune response and apoptosis. Notably, five distinct
clusters are associated with the DNA-damage response, which we were
unable to further disambiguate into more specific phenotypes.%

\paragraph{Canonical annotations.} We also color the embeddings by CORUM
protein complexes and Reactome pathways, each grouped into high-level
families. In both cases, the coloring reveals minimal spatial clustering:
genes belonging to the same complex or pathway are distributed across
the embedding space rather than forming compact groups. The exception is
the common-essential gene region, which shows partial organization by
Reactome pathway membership.

\paragraph{Interpretation.} These observations indicate that the BPMF
embedding space is organized primarily by shared screen phenotype ---
reflecting how genes co-occur as hits across historical CRISPR screens
--- rather than by canonical pathway or complex membership. This is
consistent with the finding in Section~\ref{sec:emb_init} that
embeddings which better recover canonical biological relationships
(STRING, CORUM, SIGNOR, Reactome) provide weaker initialization for
\assayformer. The BPMF embeddings capture a distinct, phenotype-derived
representation of gene function, in which proximity reflects shared
perturbation profiles across screens rather than established biological
annotations. Their strong performance suggests that these
phenotype-derived embeddings may complement existing gene embedding
approaches~\citep{littman2025gene} and motivates their evaluation in
other downstream tasks.

\FloatBarrier

\section{Biological Diversity Analysis}\label{app:diversity}

\begin{figure}[h!]
    \centering
    \includegraphics[width=1.0\textwidth]{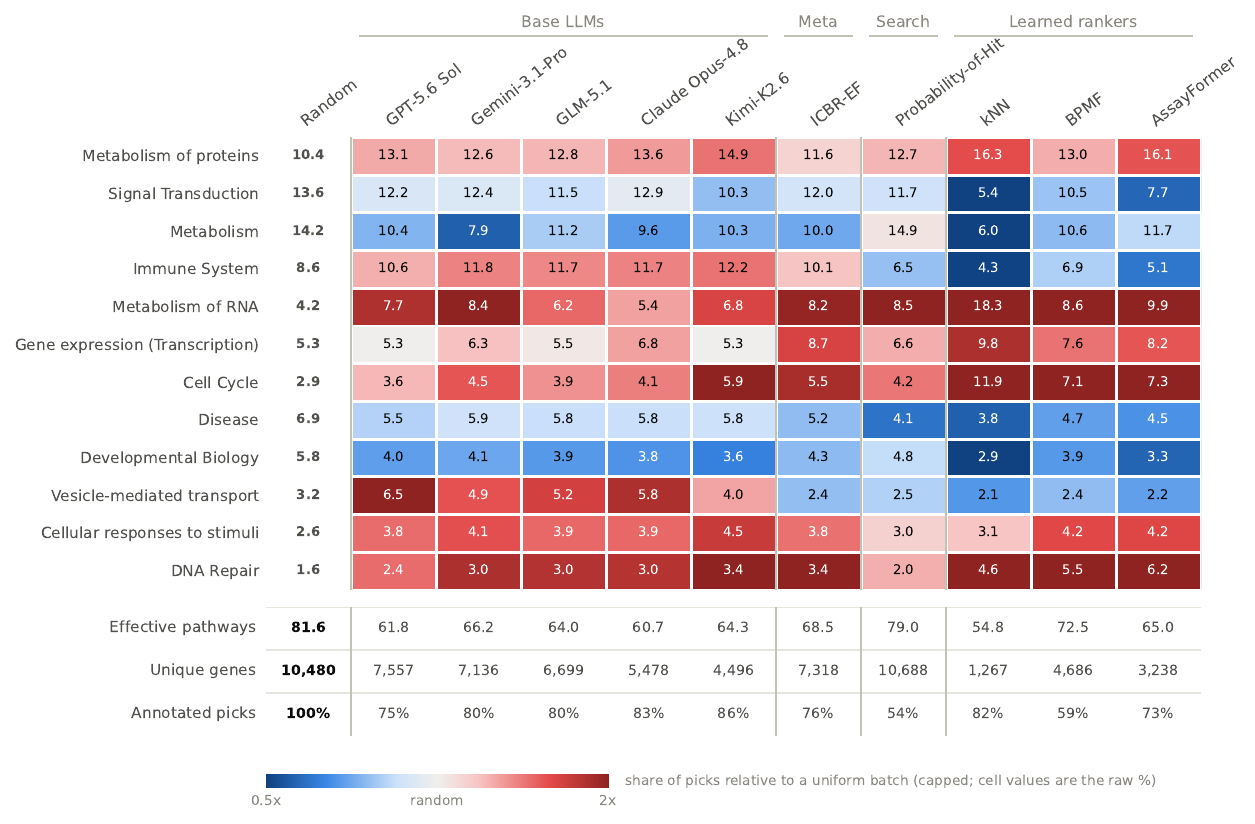}

\caption{\textbf{LLMs converge to similar distributions of biology across families.} The heatmap shows top-level composition of requested genes in terms of Reactome pathways, aggregated across the 20 test screens. Unique genes is the total number of distinct genes requested across all tested screens. Annotated picks is the percent of requested genes which have ground truth labels. The Random baseline is the pathway distribution given by uniform sampling over the universe of all possible genes. Cells are colored by by the ratio between the observed pathway share for that model and the expected share size of Random, where red = over-represented and blue = under-represented (color values capped between 0.5x and 2x). Overall, LLMs generally show similar gene acquisition signatures, with an emphasis on RNA and protein metabolism and under-representation of bulk metabolism and developmental biology. While there are differences between models, they are small compared to the difference from other baseline methodologies such as kNN. Additionally, the number of effective pathways for each LLM falls into a narrow range. Note that Kimi-K2.6 and Opus tend to supply less than the requested number of genes (100) in many cases, possibly due to biological safety training, resulting in a low number of unique genes. }

    \label{fig:diversity_heatmap}
\end{figure}

Supplementary Fig.~\ref{fig:diversity_heatmap} breaks down the Reactome composition of acquired genes by individual LLM, complementing the method-level comparison in Figure~\ref{fig:biodiversity}A.

\FloatBarrier
\newpage

\section{Metrics used in \framework}
\label{app:metrics}

\begin{figure}[htb]
    \centering
    \includegraphics[width=1.0\textwidth]{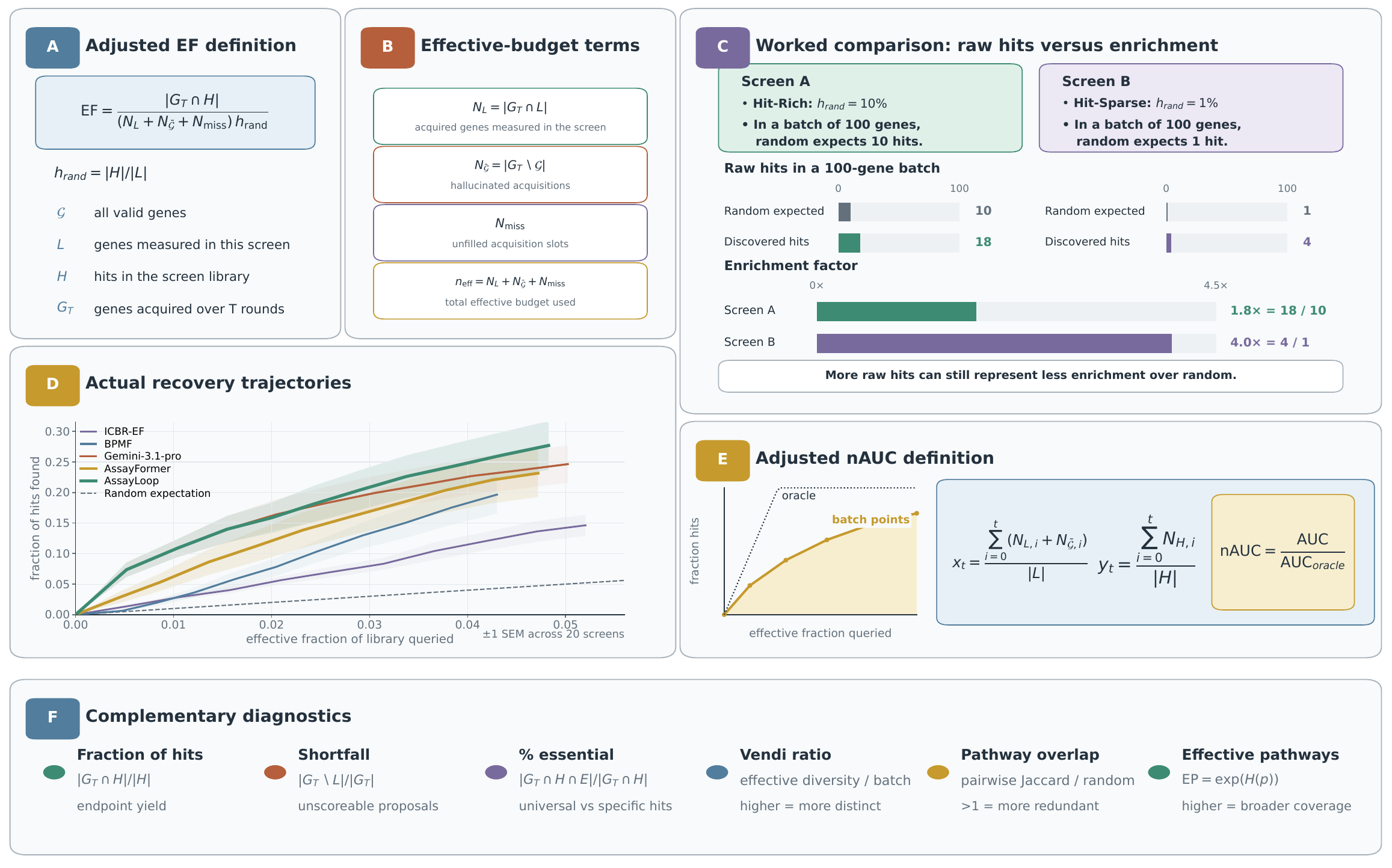}
    \caption{\textbf{Example of Enrichment Factor and other metrics}. \textbf{A)} The definition of our adjusted enrichment factor (EF). \textbf{B)} Terms used for normalizing budget calculations. \textbf{C)} A worked example of EF showing how EF is comparable across screens. Note, no adjustment for untested genes is needed in this computation. \textbf{D)} Example of hit recovery trajectories from real baselines. \textbf{E)} The definition used to calculate adjusted normalized AUC (nAUC). \textbf{F)} Additional metrics used for understanding model behavior.}
    \label{fig:metrics}
\end{figure}

Below we define the metrics used in \framework to evaluate the performance of hit discovery models. 
We let $\mathcal{G}$ denote the set of all possible genes, $L$ the set of genes in the gene library of a given screen (the set of genes for which a label $y_g$ is available), $H$ the set of hits in $L$, and $G_T$ the set of genes acquired by the model over the $T$ rounds.
Supplementary Fig.~\ref{fig:metrics} provides a visual overview of the metrics and their definitions.

\paragraph{Hit enrichment factor (EF)} Our main metric is the hit enrichment factor (EF), that measures the ratio between the number of hits found and the expected number of hits found by random selection. 

 We write $N_{L} = |G_T \cap L| $ the number of acquired genes over the full trajectory present in the screen library, $N_{\mathcal{G}} = |G_T \cap \mathcal{G}|$ the number of valid acquired genes, $N_{\bar{\mathcal{G}}} = |G_T \setminus \mathcal{G}|$ the number of hallucinated acquired genes that do not represent valid genes, and $N_{\mathrm{miss}}$ for the total
number of unfilled acquisition slots over the $T$ rounds. Finally, $h_{\text{rand}}=|H|/|L|$ denotes the random hit rate. We define the hit enrichment factor as

\begin{align}
    \label{eq:enrichment_factor_app}
    \mathrm{EF}=\frac{|G_T\cap H|}{(N_L + N_{\bar{\mathcal{G}}}+N_\text{miss})\cdot h_{\text{rand}}}.
\end{align}

This formula differs from the classical enrichment factor (\emph{i.e.,} $\frac{|G_T\cap H|}{|G_T|\cdot h_{\text{rand}}}$) in the number of acquired genes that are used for the normalization. By normalizing with $(N_L + N_{\bar{\mathcal{G}}} + N_\text{miss})$, we do not penalize acquisition of real genes ($g\in \mathcal{G})$ that are not part of the screen library but do penalize the acquisition of hallucinated genes.

\paragraph{Normalized area under the cumulative-hits curve (nAUC)}

We also report the normalized area under the cumulative-hits curve (nAUC), which captures the full acquisition trajectory rather than only the endpoint. We construct the curve by plotting the fraction of hits found against the fraction of the library effectively queried, with one point per acquisition batch. For each batch $t = 1, \ldots, T$, we denote $N_{L,t}$ and $N_{\bar{\mathcal{G}},t}$ for the cumulative number of in-library and hallucinated acquired genes in $G_t$ respectively, and $N_{H,t} = |G_t \cap H|$ for the number of hits in $G_t$. Unlike in
Eq.\ref{eq:enrichment_factor_app}, unfilled acquisition slots do not advance $x_t$, so an under-supplying
policy ends its curve at a smaller $x$ rather than being carried flat to the full budget. This keeps nAUC a measure of ordering quality. The cumulative-hits curve is then composed of points with coordinates $\big(\frac{1}{|L|} (N_{L,t} + N_{\bar{\mathcal{G}},t}), \frac{1}{|H|} N_{H,t} \big) $ for each batch.

The area under this piecewise-linear curve is computed by trapezoidal integration and normalized by the AUC of a perfect oracle that ranks all hits before any non-hit, over the same effective budget. The perfect oracle AUC is computed over a curve  with coordinates $\big(x_t,min(x_t\cdot|L|/|H|,1)\big)$, where $x_t$ are the coordinates of the predictor curve ($x_t=\frac{1}{|L|} (N_{L,t} + N_{\bar{\mathcal{G}},t})$).

\begin{align}
\label{eq:nauc}
\text{nAUC} = \frac{\text{AUC}}{\text{AUC}_{\text{oracle}}},
\end{align}
where $\text{AUC}_{\text{oracle}}$ is the area under the optimal curve evaluated at $x = (N_L + N_{\bar{\mathcal{G}}}) / |L|$. A value of $1.0$ indicates the model acquires all hits in the screen first. Random acquisition yields $\text{nAUC} \approx |H|/|L|$ (the hit rate) in expectation when $|G_T|<|H|$.

\paragraph{Fraction of hits (FH)}

We report the fraction of hits found over the whole trajectory, measuring the recall of the acquisition policy. Using the same notation, this is defined as

\begin{align}
\label{eq:frac_hits}
\text{FH} = \frac{|G_T \cap H|}{|H|},
\end{align}

i.e.\ the number of hits acquired by the model divided by the total number of hits in the screen. Only in-library genes can be hits, so out-of-library and hallucinated acquisitions do not contribute to the numerator. A value of $1.0$ means all hits in the screen were found within the acquisition budget.

\paragraph{Shortfall (SF)}

We report the shortfall as the fraction of acquired genes that fall outside the screen's gene library. Using the same notation:

\begin{align}
\label{eq:shortfall}
\text{SF} = \frac{|G_T \setminus L|}{|G_T|},
\end{align}

i.e.\ the number of acquired genes not present in the screen library divided by the total number of acquisitions. This includes both real genes outside the library ($G_T \cap (\mathcal{G} \setminus L)$) and hallucinated genes ($G_T \setminus \mathcal{G}$).

\paragraph{Percentage of essential genes ($\%_{\text{ess}}$)}

Some genes are hits in a given screen just because these genes typically lead to cell death in general, also known as essential genes. While these genes are genuine hits, they do not always reflect the particularities of the biology studied in the screen. %
We report their fraction among recovered hits as a diagnostic of how strongly acquisition is concentrated on broadly essential genes. 
We used the list of 1,827 common essential genes from the Cancer Dependency Map (DepMap)~\citep{Tsherniak2017-fa}.

Let $E$ denote the set of common essential genes, the percentage of essential genes is simply

\begin{align}
\label{eq:pct_ess}
\%_{\text{ess}} = \frac{|G_T \cap H \cap E|}{|G_T \cap H|},
\end{align}

i.e.\ the number of acquired hits that are common essentials divided by the total number of acquired hits.

\paragraph{Effective Pathways (EP)} We quantify the biological breadth of acquired genes using the effective number of Reactome level-2 pathway groups (186 groups in total). This is calculated using the first order Hill number
$$EP=\exp(H(p))$$
\noindent where $H$ is Shannon entropy and $p$ is the distribution of pathways in the sample set (assuming they were sampled uniformly). 

To make methods with different acquisition and annotation rates comparable, we rarefy each scope to a fixed number of annotated genes: \(M=30\) per batch (EP-B), \(M=200\) per screen (EP-S), and \(M=6000\) across the complete test set (EP-D). Sample sets containing fewer than the required number of annotated genes are excluded. We average over 400 Monte Carlo draws for EP-B and EP-S and 300 draws for EP-D. %
EP-B is averaged within each screen and then across screens, whereas EP-S is averaged directly across screens. 
Because a gene may be annotated to multiple groups, in each Monte Carlo draw we assign every annotated gene to exactly one of its associated groups. 
Assigning one group per gene prevents heavily annotated genes from artificially increasing diversity. 

Thus, EP is expressed as an effective number of equally represented pathway groups, with larger values indicating broader biological coverage.

\paragraph{Batch diversity metrics: Vendi score (VS) and Pathway overlap (PO)}

We report two complementary diversity metrics for the acquired batches, measuring whether the model selects functionally diverse genes or concentrates on a narrow biological neighborhood at each step.

\textbf{Vendi Ratio (VR).} For each batch, we compute gene embeddings using GenePT~\citep{Chen2024-zm}, construct the cosine-similarity Gram matrix $K$ of all acquired genes in the batch $B_t$, and compute its Vendi score~\citep{friedman2022vendi}, which corresponds to the effective number of distinct genes in $B_t$ (1 means all embeddings of genes in $B_t$ are collinear and $|B_t|$ means all embeddings are orthogonal). We report the Vendi ratio, which is normalized by $|B_t|$ and averaged across batches. A higher Vendi ratio means the genes acquired in the batch are more diverse.

\textbf{Pathway overlap (PO).} For each batch, we compute the mean pairwise Jaccard overlap of GO biological-process annotations between the acquired genes, and divide it by the same quantity for a batch of equal size drawn uniformly at random from the screen library. Averaging across batches gives PO. A value of $1.00$ corresponds to random selection, values above $1$ indicate that acquired genes share more annotations than expected by chance (a narrow biological neighborhood), and values below $1$ indicate a broader batch. PO and VS are complementary: VR measures diversity in a continuous embedding space, whereas PO measures it through discrete pathway membership.

\FloatBarrier
\newpage

\section{\assayformer Reveals Directional %
Gene-Gene Relationships}\label{app:influence}

\begin{figure}[h!]
    \centering
    \includegraphics[width=.88\textwidth]{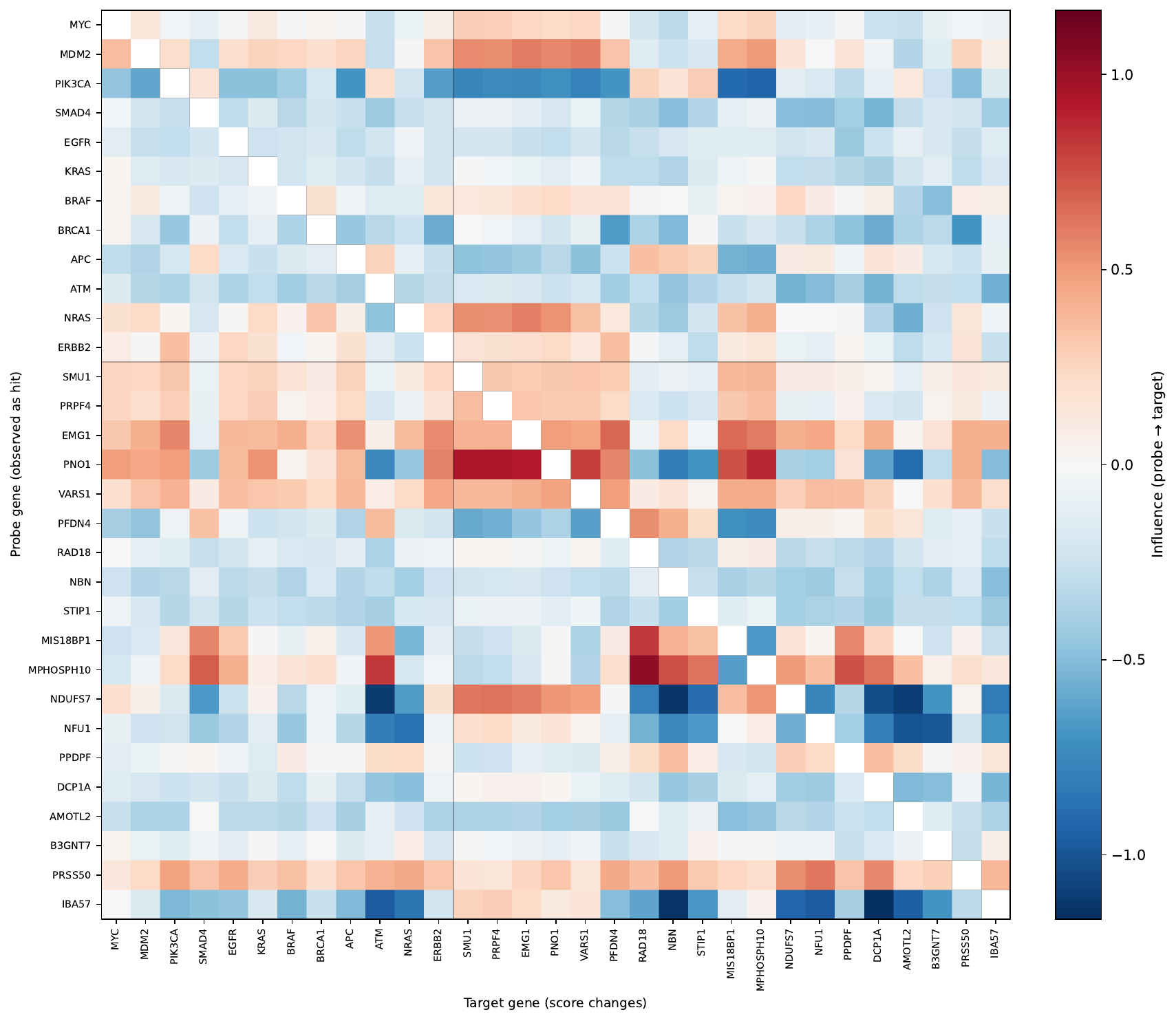}
    \caption{\textbf{Influence heatmap showing how discovered hits affect predictions.} Given a probe gene (rows) and target gene (columns), we measure the change in the acquisition score of the target gene from \assayformer when the probe gene is reported as a hit in context. The first 12 genes are important canonical cancer drivers (oncogenes and tumor suppressors). The next 19 are representative functional-module genes. }
    \label{fig:heatmap}
\end{figure}

\FloatBarrier

Because \assayformer updates its predictions as experimental
observations accumulate, we can probe the learned policy to ask how
observing one gene as a hit changes its predictions for other genes. For
a probe-target pair, we construct a random background context of 50
genes, measure the acquisition score of the target, and then measure it
again after adding the probe gene to the context as a hit. We define the
difference between these predictions as the \emph{influence} of the
probe on the target, and average this quantity across 10 independently
sampled background contexts. So we can inspect model-wide influence patterns learned by the model, we use a generic screen description ``A genome-wide CRISPR knockout screen to identify essential genes''. 

Supplementary Fig.~\ref{fig:heatmap} shows the resulting influence matrix for 31
representative genes, including 12 canonical cancer drivers and 19 genes
from functional modules. Influences among cancer drivers are relatively
weak, consistent with these genes acting through distinct biological
programs. In contrast, substantially stronger influences emerge among
functionally related gene pairs,
indicating that \assayformer has learned structured, context-dependent
relationships from historical screens.

Importantly, these relationships are strongly directional rather than
simple symmetric associations. The correlation between the influence
matrix and its transpose is only $r=0.16$, and $44\%$ of reciprocal
gene pairs have influences with opposite signs. For example, observing
MDM2 as a hit increases the acquisition score of PFDN4 by $+0.33$,
whereas observing PFDN4 as a hit decreases the acquisition score of MDM2
by $-0.46$. Thus, the model does not simply encode gene similarity:
observing a gene can induce a directional update in the predicted
relevance of another gene, and the effect need not be reciprocal.

To test whether this procedure can uncover meaningful relationships
beyond canonical annotations, we remove gene pairs represented in the
STRING, CORUM, SIGNOR, MSigDB, and Reactome databases~\citep{szklarczyk2023string,giurgiu2019corum,lo2023signor,subramanian2005gene,milacic2024reactome} and examine
high-magnitude remaining influences, shown in
Supplementary Fig.~\ref{fig:influence}. Among positively influenced pairs, we
identify relationships consistent with co-essentiality or
synthetic-lethal dependencies, such as increased spliceosome dependence
following MYC hits and increased nucleolar-stress dependence following
MDM2 hits. Conversely, negative influences highlight patterns consistent
with epistatic masking or lineage-specific exclusion, including reduced
dependence on mitotic machinery following PIK3CA hits and reduced OXPHOS
dependence following SMAD4 hits.

\begin{figure*}[t]
\centering
\includegraphics[width=\textwidth]{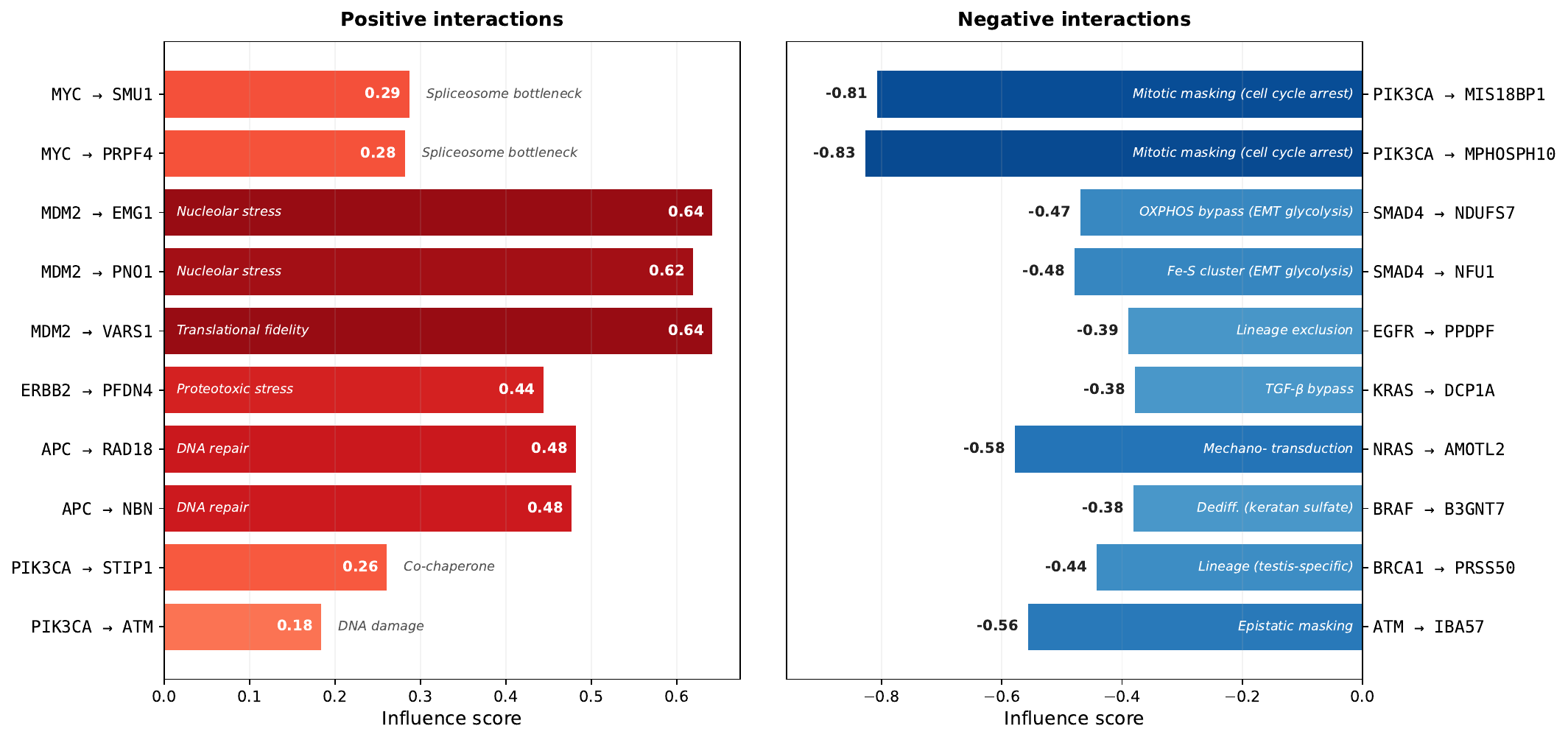}
\caption{
\textbf{Gene influence scores reveal hidden epistatic relationships not captured by standard interaction databases.}
For each probe gene (left of arrow), we measure the change in the model's predicted influence score for every target gene (right of arrow) when the probe gene is observed as a hit, averaged over random background contexts. \textbf{Left:} Boosted pairs. \textbf{Right:} Suppressed pairs. These dependencies describe the model's inferred decision process rather than established biological interactions.
}
\label{fig:influence}
\end{figure*}

\FloatBarrier

\section{Model Scaling}\label{app:model_scaling}

\begin{figure}[!ht]
    \centering
    \includegraphics[width=1.0\textwidth]{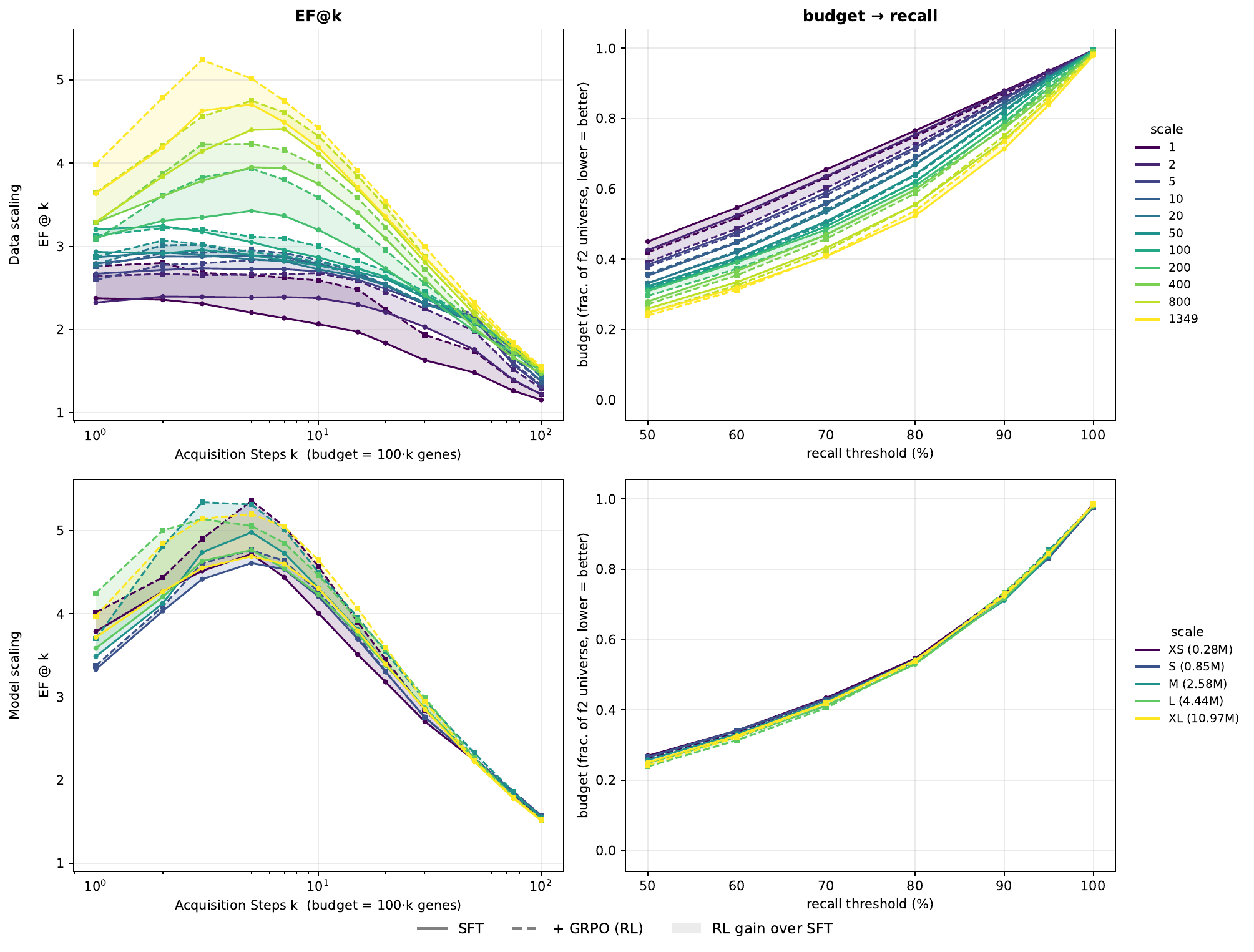}
    \vspace{-8mm}
    \caption{\textbf{Performance scales with data, but not model size.} \assayformer models (without input screen descriptions) are trained to identify scaling of in-context learning capabilities. \textbf{Top:} Models are trained on varying amounts of data from 1 to all training screens, and average performance curves are plotted. Both SFT and RL show this performance increase, with RL boosting SFT performance. \textbf{Top Left plot} is EF at $k$ acquisition steps, and shows that scaling is most effective during the first few steps of the experiment. At 100 steps (10,000) genes sampled, models approach similar performance. This is expected because even a random model will find all hits if it samples the entire genome. \textbf{Top Right plot} is budget to acquire enough hits to satisfy a recall threshold $m$. For example, to find $m=50\%$ of the hits, the `1349' model requires sampling less than 30\% of the full genome. Note that RL rollouts stop at 10 steps during training, so these curves reflect out-of-distribution gains for RL performance. \textbf{Bottom:} We also consider scaling model size from XS (0.28M parameters) to XL (10.97M parameters), given the full training set. Here, we do not observe meaningful scaling laws. We posit that larger training datasets are required to show evidence of model scaling. Note that scaling models has shown useful performance gains in pretrained LLMs on the non-active learning setting of AssayBench (Figure 4 from \citep{de2026assaybench}).}
    \label{fig:scaling}
\end{figure}

We investigate whether performance gains similar to those from additional training data can be obtained by increasing model capacity. We train 150 models spanning a nearly 40-fold range in parameter count, from 0.28M (XS) to 10.97M (XL), including our default 4.44M-parameter model (L). In contrast to data scaling, we observe no meaningful improvement with increasing model size (Supplementary Fig.~\ref{fig:scaling}). This suggests that, in the current data regime, available training screens rather than model capacity are the binding constraint on performance.

\FloatBarrier

\section{Revisiting whether LLMs can learn from lab-in-the-loop feedback}\label{app:feedback}

Recent work has investigated whether LLMs can serve directly as
acquisition policies in lab-in-the-loop biological experiments.
\citet{Gupta2025-qb} found that the performance of several LLM-based
experimental-design agents was largely insensitive to feedback provided
during the experiment, whereas \citet{Wainrib2026-rs} showed that
sufficiently capable LLMs can benefit substantially from experimental
feedback. We revisit this question on \framework by evaluating a range
of LLMs with and without access to the outcome labels from previous
acquisition rounds. Unlike these prior studies, which randomized hit
assignments in their control conditions, we simply omit the hit labels
while preserving the history of previously acquired genes. This provides
a direct measure of how much each model benefits from observing
experimental outcomes. We additionally compare to \assayformer without context; in this setting, we simply take the top 1,000 genes predicted by \assayformer with no context.

As shown in Supplementary Figure~\ref{fig:no_labels}, removing hit labels consistently
reduces performance across all tested LLMs, demonstrating that these
models do use experimental feedback to adapt their subsequent
acquisitions. This result is consistent with \citet{Wainrib2026-rs} and
confirms that adaptive in-context learning extends across multiple LLM
families. However, the magnitude of this adaptation remains limited:
standalone LLMs are consistently outperformed by \assayformer, which shows a substantially larger gain from feedback. Furthermore, \assayloop, which
utilizes an LLM for early acquisitions and delegates later acquisition rounds to \assayformer, shows both strong label-blind performance and also shows better in-context learning than LLMs. Thus, while current frontier LLMs can learn from experimental outcomes in context, a policy explicitly trained for feedback-conditioned adaptation remains more
effective for sequential hit discovery.

\begin{figure}[h!]
    \centering
    \vspace{-6mm}
    \includegraphics[width=0.65\textwidth]{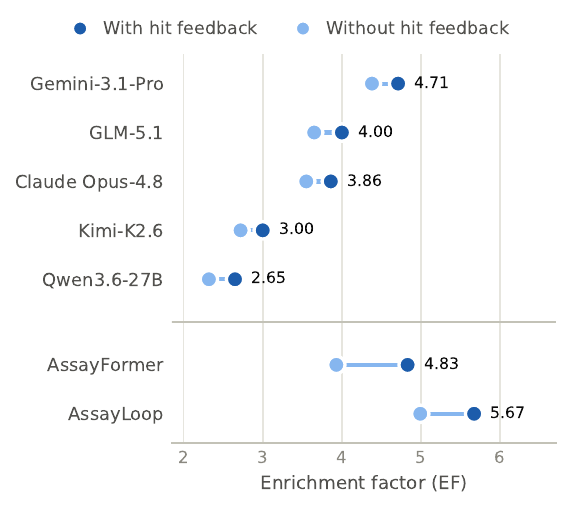}
    \vspace{-2mm}
    \caption{\textbf{LLM performance with and without per-round hit labels}. We tested the in-context learning capabilities of select LLMs by running the active learning framework \textit{without} labels from previous rounds of experiments. This allows us to distinguish between the ability of LLMs to incorporate experimental feedback with ICL, versus their warm-start capabilities without feedback. Across model families, we found that experimental feedback boosts model performance, indicating that LLMs do benefit from the active learning setting of this task. We also show the performance of \assayformer with and without context, illustrating its relatively poor ability without feedback but substantially stronger ICL capabilities. 
}
\label{fig:no_labels}
\end{figure}

\end{document}